\documentclass{article}
\usepackage{arxiv}

\usepackage[utf8]{inputenc} 
\usepackage[T1]{fontenc}    
\usepackage{hyperref}       
\usepackage{url}            
\usepackage{booktabs}       
\usepackage{amsfonts}       
\usepackage{nicefrac}       
\usepackage{microtype}      
\usepackage{lipsum}		    
\usepackage{graphicx}
\usepackage[numbers]{natbib}
\usepackage{multirow}
\usepackage[table,xcdraw]{xcolor}

\usepackage{tabularray}
\usepackage[normalem]{ulem}
\useunder{\uline}{\ul}{}

\usepackage{doi}

\title{\Large\bfseries COLLABORATION IN IMMERSIVE ENVIRONMENTS: CHALLENGES AND SOLUTIONS}

\author{ \href{https://orcid.org/0000-0002-9016-7626}{\includegraphics[scale=0.06]{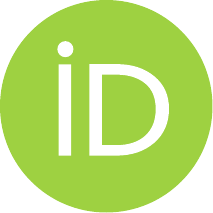}\hspace{1mm}Shahin Doroudian}\\
	Department of Computer Science\\
	University of North Carolina at Charlotte\\
	Charlotte, NC, USA \\
	\texttt{SDoroudi@uncc.edu} \\
}

\renewcommand{\shorttitle}{\textit COLLABORATION IN IMMERSIVE ENVIRONMENTS}
\hypersetup{
pdftitle={A template for the arxiv style},
pdfsubject={q-bio.NC, q-bio.QM},
pdfauthor={David S.~Hippocampus, Elias D.~Striatum},
pdfkeywords={First keyword, Second keyword, More},
}

\begin{document}

\maketitle

\begin{abstract}
Virtual Reality (VR) and Augmented Reality (AR) tools have been applied in all engineering fields in order to avoid the use of physical prototypes, to train in high-risk situations, and to interpret real or simulated results. In order to complete a shared task or assign tasks to the agents in such immersive environments, collaboration or Shared Cooperative Activities are a necessity. Collaboration in immersive environments is an emerging field of research that aims to study and enhance the ways in which people interact and work together in Virtual and Augmented Reality settings. Collaboration in immersive environments is a complex process that involves different factors such as communication, coordination, and social presence. This paper provides an overview of the current state of research on collaboration in immersive environments. It discusses the different types of immersive environments, including VR and AR, and the different forms of collaboration that can occur in these environments. The paper also highlights the challenges and limitations of collaboration in immersive environments, such as the lack of physical cues, cost and usability and the need for further research in this area. Overall, collaboration in immersive environments is a promising field with a wide range of potential applications, from education to industry, and it can benefit both individuals and groups by enhancing their ability to work together effectively.

\end{abstract}


\section{Introduction}


Since 1989 Virtual Reality has been a topic for research. Jaron Lanier \cite{10.1145/77276.77278} coined the phrase Virtual Reality for the first time in 1989 and the Commission of European Communities (CEC) identified VR as a practical technology in 1992 \cite{250922}.
Since then VR and AR tools have been applied in all engineering fields in order to avoid the use of physical prototypes, to train in high-risk situations, and to interpret real or simulated results \cite{alaker2016virtual, 9757602, 8797889}. In order to complete a shared task or assign tasks to the agents in such immersive environments, collaboration or Shared Cooperative Activities (SCA), are a necessity \cite{10.2307/2185537}.

Collaboration in immersive environments is an emerging field of research that aims to study and enhance the ways in which people interact and work together in virtual and augmented reality settings \cite{1352977, 9757558}. The increasing popularity of VR and AR technology has led to the development of new platforms and tools for collaboration, which has the potential to revolutionize how people interact and collaborate in a wide range of fields such as education, industry, healthcare and entertainment. Collaboration in immersive environments can be defined as the process of working together in a shared digital space, where users are fully immersed in the environment and can interact with one another in real-time \cite{10.1162/pres.1997.6.6.603}.

One of the key benefits of collaboration in immersive environments is the ability to create shared experiences among users \cite{10.1145/3491102.3517500}. VR and AR technology allows users to fully immerse themselves in a digital environment, which can increase feelings of social presence and improve communication and coordination among users. Additionally, immersive environments can provide users with a sense of being in the same place, even when they are physically separated \cite{10.1145/1841853.1841856}. This can be particularly useful for remote collaboration or for users who are unable to meet in person.

\begin{figure}
    \centering
    \includegraphics[width=0.4\columnwidth]{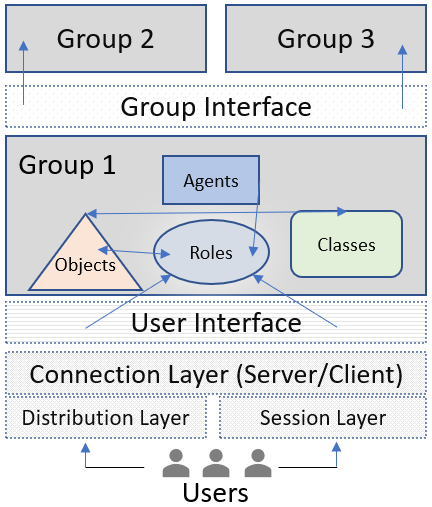}
	\caption{A Multi-user Collaborative System}
	\label{fig:fig1}
\end{figure}

Another benefit of collaboration in immersive environments is the ability to create virtual environments that are tailored to specific tasks or activities. For example, VR and AR technology can be used to create virtual training environments for medical professionals \cite{park2019literature, chheang2021collaborative, fu2022systematic}, or virtual design environments for architects \cite{6225733} and engineers \cite{8009379}. These environments can provide users with a level of realism and interactivity that is not possible with traditional methods.

Despite the potential benefits of collaboration in immersive environments, there are also challenges and limitations to this technology. One of the main challenges addressed in this paper, is the lack of physical cues, which can make it difficult for users to communicate and coordinate their actions. Additionally, usability in immersive environments can be challenging as it makes navigation and interaction difficult, particularly for non-technical users. Furthermore, there is a need for further research in this area to develop new methods and tools for collaboration in immersive environments.

Overall, collaboration in immersive environments is a promising field with a wide range of potential applications. The ability to create shared experiences and tailor virtual environments to specific tasks can enhance collaboration in various fields, from education \cite{9982707} to industry \cite{9524249}, from healthcare \cite{gerup2020augmented} to entertainment \cite{hung2021new}. However, to fully realize the potential of collaboration in immersive environments, further research is needed to address the challenges and limitations of this technology.

In this paper, we will explore the different types of immersive environments and the different forms of collaboration that can occur in these environments. We have consolidated the research on immersive VR/AR environments to address the following research questions: 1) How can domain experts and organizations benefit from Immersive Collaboration in AR/VR and 2) what are the main research areas for Immersive collaboration? We will also discuss the challenges and limitations of collaboration in immersive environments, such as the lack of physical cues, cybersickness and latency. Additionally, we will examine the potential benefits of collaboration in immersive environments and the ways in which this technology can be used to enhance collaboration in various fields. The paper is organized as follows: In section 2 we will discuss the current existing challenges in collaboration in immersive environments. In section 3 we will cover the methodologies and talk about how these challenges are addressed. Later in section 4 various applications are reviewed for a better perspective on this topic.

\section{Challenges}

As briefly discussed in the previous section of this discourse, while the potential advantages of collaboration in immersive environments are numerous, there are also several critical challenges that need to be addressed in order to fully realize the benefits of this technology \cite{10.1145/3313831.3376722}. In this section, we will delve more deeply into these challenges, examining them in detail. Through this analysis, we will gain a better understanding of the obstacles that can hinder the effectiveness and usability of immersive collaboration tools, and how we can design and develop them to achieve optimal outcomes. By carefully considering these challenges and developing targeted solutions, we can create immersive collaboration experiences that are more accessible, user-friendly, and more effective for all users.

\begin{figure}[h]
    \centering
    \includegraphics[width=0.5\columnwidth]{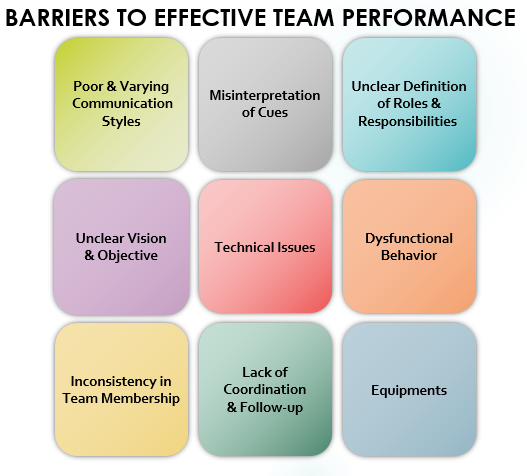}
	\caption{Performance Components in a Collaborative Domain.}
	\label{fig:fig0}
\end{figure}


\subsection{Skill level of team members}
The skill level of team members is an important factor to consider when examining the role of leader/facilitator presence and behavior on collaborative work outcomes in immersive environments \cite{digirolamo2019exploration}. Not all team members may be equally familiar or comfortable with the technology used in these environments, which can affect their ability to collaborate effectively.

For example, in a virtual reality environment, team members may need to have a certain level of familiarity with the technology to navigate the space and interact with objects. If some team members lack this skill, they may require additional support and guidance from the leader/facilitator, which can be challenging in a virtual setting. Similarly, in an augmented reality environment, team members may need to be familiar with the physical space and how to interact with objects and digital overlays. If some team members lack this skill, they may struggle to effectively collaborate with others in the environment.

In both cases, the leader/facilitator may need to provide additional support and guidance to team members who are less familiar with the technology or the physical space. This can be particularly challenging in immersive environments, where communication and coordination may be more difficult than in person.

Effective leaders/facilitators must be aware of the skill level of team members and be prepared to provide additional support and guidance as needed to ensure that everyone is able to effectively collaborate towards a common goal.

\subsection{Communication barriers}
Communication barriers are one of the main challenges that can arise when working in immersive environments, and can have a significant impact on collaborative work outcomes \cite{jenifer2015cross}. In immersive environments, team members may not be able to communicate as effectively as they would in person, which can lead to misunderstandings, delays, or errors.

One of the main communication barriers in immersive environments is the lack of nonverbal cues \cite{huang2021interactivity}. Nonverbal cues such as facial expressions, body language, and tone of voice can provide important information about a person's emotions or intentions, and help to establish rapport and trust between team members. In immersive environments, these cues may be more difficult to detect or may not be present at all, which can lead to misinterpretations or misunderstandings.

Another communication barrier in immersive environments is the reliance on technology. Immersive environments often rely on complex technology, which can be prone to malfunctions or may not always be readily available. Technical difficulties can interrupt communication and collaboration, leading to delays and frustration.

Finally, language barriers can also pose a communication challenge in immersive environments. Collaborative work often involves individuals from different cultural and linguistic backgrounds, and the leader/facilitator must be aware of any language barriers that may affect communication and collaboration.

Overall, communication barriers can have a significant impact on collaborative work outcomes in immersive environments. Effective leaders/facilitators must be aware of these challenges and work to mitigate them by promoting clear and open communication, encouraging feedback, and providing training and support for team members to improve their communication skills.

\subsection{Cultural differences}
Cultural differences are another important factor to consider when examining the role of leader/facilitator presence and behavior on collaborative work outcomes in immersive environments. Collaborative work often involves individuals from different cultural backgrounds, and these cultural differences can affect communication and collaboration in immersive environments.

For example, cultural differences in communication styles may affect how team members express themselves, or how they interpret others' messages \cite{tannen1983cross}. In some cultures, direct communication may be valued, while in others, indirect communication may be preferred \cite{colleoni2013csr}. These differences can lead to misunderstandings or misinterpretations in collaborative work, particularly in immersive environments where nonverbal cues may be limited.

Cultural differences in work styles can also affect collaboration in immersive environments. For example, in some cultures, individual achievement may be emphasized over teamwork, while in others, group harmony may be prioritized over individual goals. These differences can affect how team members approach collaborative work, and may require the leader/facilitator to adjust their behavior to accommodate these differences.

Finally, cultural differences in decision-making styles can also affect collaboration in immersive environments. For example, some cultures may prefer a consensus-based approach to decision-making, while others may prioritize quick decision-making by a designated leader. These differences can affect the speed and effectiveness of decision-making in collaborative work, particularly in immersive environments where communication may be more difficult.

Overall, cultural differences can pose a significant challenge to collaboration in immersive environments. Effective leaders/facilitators must be aware of these differences and work to promote a collaborative environment that accommodates different communication styles, work styles, and decision-making styles. This may require additional training and support for team members, as well as a willingness to adapt to the unique challenges of working in immersive environments.

\subsection{Technical limitations}

Despite the emerging advancements in the field of virtual reality, this technology is still in its early stages, and there may be inherent limitations in terms of graphics, processing power, and connectivity \cite{van2010survey} that can significantly impact the feasibility and efficacy of collaboration in virtual reality environments. As VR collaboration tools become more sophisticated and widely adopted, it is essential to recognize the potential limitations of this technology and to develop targeted solutions that can address these issues. This requires a comprehensive understanding of the technical and practical considerations involved in designing and deploying immersive collaboration tools, as well as a commitment to ongoing innovation and improvement to meet the evolving needs of users and organizations. By proactively addressing these limitations and leveraging emerging technologies and design principles, we can build more robust, accessible, and effective VR collaboration tools that can drive enhanced collaboration, innovation, and productivity in a range of contexts.

\begin{figure}[ht]
    \centering
    \includegraphics[width=0.8\columnwidth]{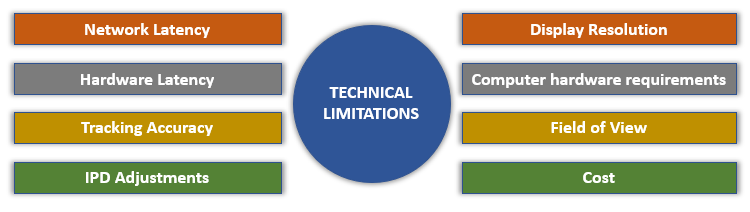}
	\caption{Technical limitations that can impact the ability of the user to effectively guide and support team members in immersive collaborative work environments.}
	\label{fig:fig}
\end{figure}

\subsubsection{Network Latency}
Immersive collaboration systems often rely on a high-speed, low-latency network connection to provide real-time collaboration between multiple users \cite{8743304}. This can be a challenge in areas with limited network infrastructure or in situations where users have limited bandwidth.
Network latency is a significant challenge in immersive environments, as it can greatly impact the overall user experience.
It also poses some challenges that may hinder the learning experience. For instance, technical issues like poor internet connectivity or device compatibility can impede the effectiveness of remote learning. Similarly, a lack of interaction and socialization, which is prevalent in traditional classroom settings, may make remote learners feel isolated and disengaged from their studies. This can lead to a lack of motivation, which may further compound the challenges of remote learning \cite{pellas2019augmenting}. Furthermore, assessing learners' progress and ensuring academic integrity in a remote learning environment can be challenging, as it is more difficult to monitor their activities and prevent cheating.

Immersive environments typically require real-time, low-latency communication between multiple devices in order to create a seamless and convincing virtual world. However, if there is a high amount of latency in the network connection, the virtual environment can become unresponsive, jittery, or otherwise disorienting for the user \cite{9416950}.

This is especially true in multi-user immersive environments, where multiple users are interacting in the same virtual space. Network latency can lead to delays in transmitting user inputs and updates \cite{9757480}, which can result in a lack of synchronization between users and a feeling of disconnection.

To mitigate the impact of network latency in immersive environments, it is important to have a robust and fast network infrastructure, as well as efficient communication protocols and compression algorithms \cite{9913710}. Some systems may also implement prediction algorithms or other techniques to minimize the impact of latency on the user experience.

\subsubsection{Hardware Latency}

The lag between head movement and the corresponding change in the VR environment can cause discomfort, known as cybersickness \cite{stauffert2020latency}. Hardware latency, or the time it takes for a device to respond to an input, can be a significant limitation in collaboration in immersive environments \cite{9268953}. Latency can affect the user's experience in several ways:

\textbf{Slow Responsiveness:} High latency can make the virtual environment feel slow and unresponsive, which can be frustrating for users and impact their ability to collaborate effectively.

\textbf{Disruptive Experience:} Latency can create a noticeable delay between the user's actions and the response in the virtual environment, disrupting the flow of the experience and making it difficult for users to collaborate seamlessly.

\textbf{Inconsistent Experience:} Latency can vary depending on the user's hardware and network conditions, leading to an inconsistent experience for different users. This can make it difficult for users to coordinate their actions and work together effectively.

\textbf{Increased Error:} Latency can also increase the likelihood of errors, as users may make incorrect actions due to a delay in the response time.

In conclusion, hardware latency can be a significant limitation in collaboration in immersive environments, as it can impact the user's ability to interact with the virtual environment in a responsive and seamless manner. It is important to consider latency when designing immersive environments for collaboration, and to use high-performance hardware and low-latency networks to minimize the impact of latency on the user experience.

\subsubsection{Tracking Accuracy}
Tracking accuracy in virtual environments refers to the ability of a tracking system to accurately and precisely determine the position and orientation of an object or user within the virtual environment.

There are several factors that can impact tracking accuracy in virtual environments, including the quality of the tracking system, the calibration of the system, the type of markers or sensors being used, and the complexity of the environment \cite{10.1145/3485279.3485283}.

To improve tracking accuracy in virtual environments, it is important to use a high-quality tracking system that is properly calibrated and configured for the specific use case \cite{pastel2021comparison}. This may involve using multiple sensors or markers to track movement, or using advanced algorithms to filter out noise and improve the accuracy of the data.

It is also important to consider the environment in which the tracking is taking place. In some cases, it may be necessary to create a dedicated space for tracking that is free from interference or distractions. Additionally, it may be necessary to adjust the lighting conditions or other factors in the environment to ensure that the tracking system is able to accurately capture data.

Overall, achieving high tracking accuracy in virtual environments requires careful attention to detail and a commitment to using the best available technology and techniques to capture and analyze data.
VR systems need precise tracking of the user's head and hand movements to provide an immersive experience, and small tracking errors can result in an unconvincing experience.

\subsubsection{Interpupillary Distance Adjustments}

Different people have different IPDs, and VR systems need to be adjusted accordingly for a comfortable and effective experience \cite{livitcuka2023impact}. IPD adjustments in virtual environments refer to the process of calibrating the distance between the lenses or screens of a virtual reality headset to match the distance between a user's pupils. This adjustment is important to ensure that the virtual environment appears correctly scaled and in focus, and to minimize any discomfort or eye strain that may result from using a headset with an incorrect IPD setting.

Adjusting the IPD of a virtual reality headset typically involves using software controls to modify the distance between the two lenses or screens \cite{9376369}. Some headsets may also feature physical adjustments that allow users to manually adjust the distance between the lenses.

To determine the correct IPD setting for a user, it is typically necessary to measure the distance between the user's pupils. This can be done using a specialized tool or app that uses the camera on a mobile device to capture an image of the user's face and determine the correct IPD setting based on the measurements of their facial features.

Once the correct IPD setting has been determined, it is important to ensure that the headset is properly calibrated and configured to use that setting. This may involve adjusting software settings, physical adjustments to the headset, or both.

Overall, adjusting the IPD in virtual environments is an important step in ensuring a comfortable and immersive experience for users, and can help to minimize discomfort and eye strain that may result from using a headset with an incorrect IPD setting.

IPD adjustments are also important in collaborative virtual environments, where multiple users are interacting with each other in the same virtual space. In these environments, it is important to ensure that each user's perspective is properly calibrated to match their individual IPD, as this can help to improve the sense of presence and immersion for each user.

>>Solution>>There are several ways to approach IPD adjustments in collaborative virtual environments. One common approach is to use software tools that allow each user to input their own IPD settings, which are then used to adjust the virtual environment to match. This can be done either before the collaborative session begins, or dynamically during the session as new users join and their IPD settings are detected.

Another approach is to use specialized hardware that can dynamically adjust the IPD settings for each user in real-time. For example, some virtual reality headsets feature eye-tracking technology that can detect the position of each user's eyes and automatically adjust the IPD settings to match. This can help to ensure that each user's perspective is properly calibrated without the need for manual adjustments.

Overall, IPD adjustments are an important consideration in collaborative virtual environments, as they can help to improve the sense of presence and immersion for each user, and ensure that the virtual environment is properly scaled and in focus for all participants. By using software and hardware tools that are designed to support IPD adjustments, it is possible to create collaborative virtual environments that are comfortable, engaging, and immersive for all participants.

\subsubsection{Display Resolution}
VR displays still lack the pixel density needed to eliminate the "screen door effect" where individual pixels can be seen \cite{8797975}. Graphics quality is a critical factor in creating a compelling and believable immersive experience. In order for an immersive environment to be effective, it must have high-quality graphics that are smooth, detailed, and responsive to user actions. Poor graphics quality can detract from the user experience and make it difficult for the user to fully immerse themselves in the virtual environment.
Some of the technical challenges in achieving high graphics quality in immersive environments include:

\textbf{Processing Power:} Creating high-quality graphics in real-time requires a significant amount of computational power. This can be a limitation for older computers or mobile devices, and can also impact the performance and battery life of the device \cite{feng2022wayfinding}.

\textbf{Graphics APIs:} Immersive environments typically require specialized graphics APIs, such as WebGL or Vulkan, in order to create high-quality graphics. These APIs can be complex to use and may require significant development resources to implement \cite{ferraz2021benchmarking}.

\textbf{Graphics Assets:} Creating high-quality graphics assets, such as 3D models and textures, can be a time-consuming and resource-intensive process. This can be a significant challenge for developers, especially in environments where the graphics must be highly detailed and interactive \cite{gattullo2020informing}.

\textbf{Graphics Optimization:} Optimizing graphics for performance can be a complex and ongoing process. This can include reducing the number of polygs in 3D models, using efficient lighting algorithms, and reducing the number of draw calls.

By developing innovative solutions to overcome the aforementioned technical challenges such as addressing processing power limitations, utilizing specialized graphics APIs, streamlining graphics asset creation and applying effective graphics optimization techniques, developers can successfully create immersive environments with high graphics quality that provide a believable and engaging experience for the user, thereby creating an enhanced level of user experience that can lead to greater user engagement and satisfaction.

\subsubsection{Computer hardware requirements}
High-end VR systems require a powerful computer to run smoothly and deliver a convincing virtual experience \cite{delgado2020research}. The hardware requirements for collaboration in virtual environments depend on a variety of factors, including the complexity of the virtual environment, the number of users participating in the collaboration, and the types of interactions that are taking place.

In general, collaborative virtual environments require more processing power and graphics capabilities than traditional applications \cite{de2020survey}, as they must render 3D graphics and track multiple users in real-time. As a result, it is important to ensure that the computer hardware used to access the virtual environment is capable of handling the demands of the application.

The specific hardware requirements for collaboration in virtual environments can vary depending on the platform being used. Some virtual reality platforms, for example, require high-end gaming computers with powerful graphics cards and fast processors. Other platforms may be more lightweight and can run on a wider range of hardware configurations.

In addition to the hardware requirements for accessing the virtual environment, it may also be necessary to have specialized input devices, such as VR controllers or haptic feedback devices, to facilitate collaboration and interaction within the virtual environment.

Overall, the hardware requirements for collaboration in virtual environments can be significant, and it is important to carefully evaluate the needs of the application and the capabilities of the hardware before beginning a collaboration. By choosing hardware that is capable of handling the demands of the virtual environment, it is possible to create a smooth, engaging, and immersive collaborative experience for all participants.

\subsubsection{Field of View}
The Field of View (FOV) is the extent of the observable environment visible to the user in a virtual environment. The FOV of current VR headsets is limited, which can detract from the sense of presence in virtual environments \cite{10.1145/2543581.2543590}. A limited FOV can make the user feel like they are looking through a narrow window rather than being fully immersed in the virtual environment. This can reduce the user's ability to feel like they are really there and limit their effectiveness in collaborating with others \cite{masnadi2022effects}.

A narrow FOV can also limit the user's ability to see and interact with the entire virtual space, reducing their effectiveness in collaborating with others. For example, in a virtual meeting, if a user cannot see all the participants or the shared virtual workspace, it can make it difficult for them to fully engage in the meeting and contribute to the discussion. This can ultimately hinder the effectiveness of the collaboration \cite{trepkowski2019effect}. Furthermore, if different users have different FOVs, it can create a disconnect between what each person sees, making it difficult for them to coordinate their actions and collaborate effectively. For example, if one user can see a virtual object that another user cannot, it can lead to miscommunications and misunderstandings. This can ultimately hinder the effectiveness of the collaboration and reduce the quality of the final outcome.

The level of immersion in the virtual environment can also be reduced with a limited FOV, which can hinder the user's ability to fully engage in the experience and collaborate with others. Immersion refers to the feeling of being fully absorbed and engaged in a virtual environment. A limited FOV can reduce the level of immersion, making the virtual environment feel less real and reducing the user's motivation to engage fully in the experience.

Moreover, a narrow FOV can cause frequent interruptions, as the user is forced to constantly adjust their view in order to see different parts of the virtual space. This can lead to a disjointed experience, causing the user to become disoriented and reducing their effectiveness in collaborating with others \cite{wibowo2021improving}.

Overall, a limited FOV can have a significant impact on the effectiveness of collaboration in immersive environments, reducing the user's ability to fully engage in the experience and collaborate effectively with others. It is important for developers to continue to innovate and improve the FOV of VR headsets to enhance the user's experience and facilitate collaboration in virtual environments.




\subsubsection{Cost}

VR technology can be expensive, and the cost of VR hardware and software remains a barrier to wider adoption of the technology.  The cost of hardware for immersive environments, such as VR and AR systems, can vary widely depending on the type and quality of the hardware being used \cite{farra2019comparative}.

For VR systems, the cost can range from a few hundred dollars for entry-level systems such as the Oculus Quest 2, to several thousand dollars for high-end systems such as the Valve Index or the HP Reverb G2. The cost of VR hardware can be influenced by factors such as display resolution, tracking capabilities, and the quality of the controllers and other accessories.

For AR systems, the cost can also vary widely depending on the type of system being used. Mobile AR systems, which use a smartphone or tablet to overlay digital content onto the real world, are generally less expensive than dedicated AR headsets such as the Microsoft HoloLens or the Magic Leap One. The cost of AR hardware can be influenced by factors such as display quality, tracking capabilities, and the level of integration with other systems and software.

In addition to the cost of the hardware itself, it is also important to consider the cost of any additional accessories or software that may be required to use the hardware effectively. This can include items such as specialized controllers, tracking systems, and software development tools \cite{dincelli2022immersive}. High-end VR software can be expensive, limiting its accessibility for some users. When considering VR collaboration, it's important to understand that the cost of the device is not the only expense to take into account. Additional costs may include software, and ongoing maintenance and support, which can add up quickly. Moreover, if you plan to collaborate with others in different locations, you may also need to consider the cost of a network infrastructure to support the collaboration. Aside from the cost, there are other important factors to consider when evaluating VR collaboration applications. These include ease of use, platform compatibility, security, and support and maintenance requirements. Additionally, VR collaboration may require additional hardware and infrastructure, such as VR headsets and high-speed internet connections, which can also impact the overall cost. The cost of VR collaboration can vary widely depending on several factors such as the complexity of the project, the technology used, and the scale of the collaboration. To make an informed decision, it's essential to consider all of these factors and determine which VR collaboration solution best meets your needs and budget. The cost of VR collaboration can be significant, but the benefits of enhanced collaboration and engagement may outweigh the costs. Organizations considering VR collaboration should carefully evaluate the costs and benefits of the project to determine whether it is a worthwhile investment.




Overall, the cost of hardware for immersive environments can range from a few hundred dollars to several thousand dollars, depending on the specific needs and requirements of the user. By carefully evaluating the capabilities and costs of different hardware options, it is possible to choose a system that is both effective and affordable.

\subsection{Interoperability}
Different VR platforms and devices can be incompatible, making it difficult for users to collaborate and share experiences. Interoperability in VR refers to the ability of VR systems, devices, and content to work seamlessly with each other \cite{dris2019openbim, shirowzhan2020bim}. Currently, there are several barriers to interoperability in VR, including:

\textbf{Proprietary standards:} Different VR manufacturers use their own proprietary standards and technologies, which can prevent devices from different brands from working together.

\textbf{Content compatibility:} VR content developed for one platform may not be compatible with other VR systems, limiting the ability of users to access and enjoy VR experiences on different devices.

\textbf{Input device compatibility:} Different VR systems may use different input devices, such as motion controllers or hand tracking, which can limit the ability of users to interact with VR content across different devices.

To address these barriers and improve interoperability in VR, there have been efforts to establish industry-wide standards and open-source initiatives that aim to promote compatibility and compatibility across different VR systems. By improving interoperability, VR technology can become more accessible and easier to use for consumers, leading to wider adoption and a more vibrant VR ecosystem.

\subsection{Cybersickness}

Motion sickness induced by virtual reality can cause a variety of autonomic responses in the body. The symptoms of motion sickness can include nausea, dizziness, sweating, pallor, and increased salivation, among others \cite{ohyama2007autonomic, 10.1145/3411764.3445701, CHEN201517}. These symptoms are caused by a mismatch between the visual and vestibular (inner ear) inputs, which can confuse the brain and lead to a feeling of motion sickness.

The autonomic responses to motion sickness induced by virtual reality are mediated by the sympathetic and parasympathetic nervous systems. The sympathetic nervous system is responsible for the "fight or flight" response \cite{6496840}, which can lead to increased heart rate, sweating, and other responses that prepare the body for action. The parasympathetic nervous system is responsible for the "rest and digest" response, which can slow heart rate, increase digestion, and relax the body.

motion sickness including eye fatigue, headaches, nausea, and sweating
High latency or lag can cause motion sickness or discomfort for users, impacting the quality of the VR experience.
-
Overall, the autonomic responses during motion sickness induced by virtual reality can be uncomfortable and unpleasant for users. To reduce the incidence of motion sickness in virtual reality, designers and developers can use various techniques, such as reducing the field of view, reducing latency, and incorporating natural movement into the virtual environment, to minimize the sensory mismatch that causes motion sickness. Additionally, users can take breaks and acclimate themselves gradually to the virtual environment to reduce the likelihood of motion sickness.

 Cybersickness is a type of motion sickness that can occur when using virtual reality (VR) technology. It is caused by the disconnect between what the user sees and feels in the virtual environment and what their body is actually experiencing. Symptoms include dizziness, nausea, headaches, and eye strain. Cybersickness can be reduced by incorporating physical cues into VR experiences, limiting VR session lengths, and allowing for breaks in between VR sessions.

\subsection{Social presence}

Social presence in virtual reality refers to the feeling of being present with other people in a virtual environment, as if they were physically there. Social presence is an important aspect of virtual reality because it affects how users feel about the experience and how engaged they are with the virtual world and other users. Collaboration in immersive environments can present challenges to social presence, particularly when users are physically separated. Several factors contribute to social presence in virtual reality. One such challenge is the limitation of nonverbal cues like facial expressions, body language, and eye contact \cite{lee2021kinect, balbin2019augmented}, which are crucial for effective communication and collaboration. Physical cues in virtual reality refer to sensory stimuli that mimic the physical sensations of the real world in a virtual environment and enhance the user's immersion to make the virtual environment feel more real \cite{nijholt2022capturing}. These can include haptic feedback through wearable devices, wind simulations, heat or cold sensations through temperature-controlled surfaces, vibrations, pressure, texture and more. The purpose of these cues is to enhance the sense of presence and immersion in the virtual world, making the experience more realistic and engaging. These limitations can result in misunderstandings and a lack of social presence. Another challenge in immersive environments is inconsistent or delayed feedback, which can hinder effective collaboration and reduce social presence \cite{voit2019online}. Technical issues, as mentioned in the previous sections, such as latency, bandwidth limitations, and hardware malfunctions can also disrupt communication and collaboration, resulting in reduced productivity and social presence. Cultural differences should also be considered as a potential challenge in diverse teams, as differences in communication styles, values, and norms can lead to misunderstandings and a lack of social presence. Finally, trust is critical for effective collaboration, and a lack of trust can hinder teamwork, productivity, and social presence.

By taking into account these challenges and identifying effective solutions, designers and developers can create immersive and engaging collaborative environments that maximize social presence and facilitate effective teamwork and productivity, even when team members are not physically together. For example, the use of advanced audio and video technologies can enhance communication and create a more immersive experience, while the implementation of real-time feedback mechanisms can facilitate quick and effective responses to feedback. Additionally, fostering a culture of trust and openness can help to build strong relationships among team members and facilitate effective collaboration. By addressing these challenges and leveraging appropriate tools and strategies, collaborative environments can be optimized to promote social presence, teamwork, and productivity.

\subsection{Usability}
Navigating and interacting with virtual environments can be a daunting task, especially for non-technical users who may not be familiar with the interface and the commands required to operate in the virtual environment \cite{diersch2019potential}. These users may find it challenging to complete even the most basic tasks, such as moving around or selecting objects within the virtual environment. Additionally, virtual environments require specific hardware and software configurations, which non-technical users may not possess or may not be able to set up on their own. This can make it difficult for them to access and use the virtual environment, leaving them feeling excluded and disconnected from the technology. 

Furthermore, virtual environments can be complex and overwhelming \cite{hejtmanek2020much}, which can make it even more challenging for non-technical users to understand and use them effectively. They may require a significant amount of time and effort to master, and without the necessary guidance and support, users may become disengaged and uninterested in the technology. This lack of engagement can result in poor adoption rates and a lack of interest in exploring the possibilities offered by virtual environments. It is essential to provide adequate training and support to non-technical users to help them feel more confident and comfortable using virtual environments \cite{10.1145/3343036.3343131}. This could include resources such as video tutorials, user guides, or interactive training modules to help users develop the skills they need to interact with the virtual environment successfully. By addressing these challenges and providing appropriate support, designers and developers can make virtual environments more accessible, engaging, and useful for all users, regardless of their technical background or experience level.
Overall, the use of virtual environments can be incredibly beneficial, but the challenges of navigating and interacting with them can be particularly daunting for non-technical users. However, providing appropriate guidance and support can help users overcome these obstacles and fully realize the potential of the technology. By offering detailed instructions, user guides, and training materials, designers and developers can help users feel more confident and comfortable using virtual environments. Additionally, the development of more intuitive and user-friendly interfaces can help to reduce the learning curve and make it easier for non-technical users to interact with the virtual environment \cite{yang2019gesture}.

Effective support and guidance can not only help users feel more comfortable using virtual environments, but it can also enhance their engagement with the technology. Users who feel supported and empowered to use the technology are more likely to explore its various features and capabilities, leading to increased productivity and improved outcomes. By taking steps to address the challenges that non-technical users face when interacting with virtual environments and providing the necessary support, designers and developers can ensure that the technology is accessible and beneficial to everyone, regardless of their technical expertise or experience level.

\subsection{Data privacy and security}

As VR collaboration becomes increasingly prevalent, there is a growing concern regarding data privacy and security. The potential for sensitive information to be shared within virtual environments raises questions around the collection, use, and storage of user data \cite{10.1145/3359626}. These concerns have been amplified as more individuals and organizations adopt VR technology for various purposes, including collaboration and communication \cite{jozani2020privacy, ali2018privacy}.

VR applications and platforms have the potential to collect a wide range of user data, including personal information, location data, and behavioral data. To ensure that data privacy is maintained, it's crucial to understand what data is being collected, why it's being collected, and how it's being used and stored. Users should have the ability to control their data and opt-out of data collection if they choose to do so. To further protect sensitive data and virtual environments, these platforms should have robust authentication and access control mechanisms. This includes implementing strong passwords, multi-factor authentication, and user roles and permissions.

\begin{figure}[ht]
    \centering
    \includegraphics[width=0.6\columnwidth]{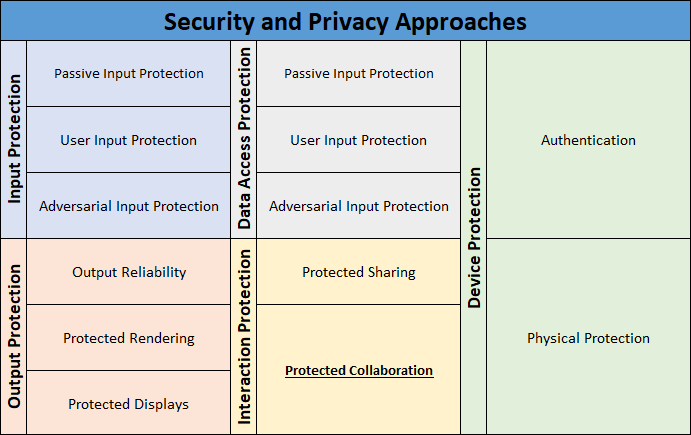}
	\caption{Data Privacy and Security Classification. You can see Collaboration is classified under Interaction Protection section}
	\label{fig:fig3}
\end{figure}

Furthermore, to ensure that sensitive information remains secure, VR applications and platforms should use secure protocols such as HTTPS and SSL to ensure that data transmitted between users and servers is encrypted and secure. Clear policies and procedures for data storage and retention should also be established to ensure that sensitive data is stored securely and only retained for as long as necessary. Compliance with relevant data privacy and security regulations, such as GDPR and CCPA, is also essential \cite{harding2019understanding}.

In today's digital landscape, where data breaches and cyber attacks are becoming increasingly common, it is crucial for both users and organizations to be aware of the potential privacy and security risks associated with VR technology. This includes understanding the types of data that may be collected, how it may be used, and the potential consequences of a data breach or cyber attack.

To mitigate these risks, users should carefully review the privacy and security policies of VR applications and platforms. This involves understanding what information is being collected, how it is being used, and whether it is shared with third parties. Users should also be cautious about sharing personal information within virtual environments and limit the amount of data they provide to these platforms.

In the following section of this discussion, we will delve into the methodology employed in virtual reality collaboration and examine some of the feasible solutions that have been proposed to address the challenges identified earlier. This exploration will highlight how designers and developers are leveraging innovative technologies and design principles to improve the accessibility, user-friendliness, and efficacy of virtual reality collaboration tools. We will also examine how these solutions can be adapted to specific use cases and contexts, in order to achieve optimal outcomes and benefits for different types of users and organizations.

\section{Methodology}
Collaboration in immersive environments involves working together in virtual reality or augmented reality environments to achieve a common goal. There are several methodologies for collaboration in immersive environments that can help to ensure effective teamwork and positive outcomes. Here are some of the key methodologies for collaboration in immersive environments:
%

\subsection{Networking}

There are several solutions for VR collaboration networking that can help to address the challenges of bandwidth, latency, and reliability that can impact the effectiveness of virtual reality collaboration. The development of 6G technology is expected to revolutionize the future of VR collaboration networking \cite{10070393}. An important aspect of 6G technology is its potential to enable more advanced collaboration in AR and VR. With improved data speeds and reduced latency, it is expected that multiple users can collaborate more effectively in the same virtual environment. This can have significant implications for industries such as education, training, and remote work, where collaborative AR and VR environments can enhance productivity and learning outcomes \cite{9656726, rashvand2024realtimebusarrivalprediction}. Furthermore, 6G technology can also facilitate the development of new AR and VR applications, such as smart cities and autonomous vehicles, which require high-speed data transfer and low latency connections \cite{giordani2020toward}. These applications can have significant social and economic benefits, such as improved transportation efficiency and enhanced urban planning. Several other solutions can optimize networking for VR collaboration, including cloud-based infrastructure, dedicated high-speed networks, edge computing, network optimization techniques, and Quality of Service (QoS) or Quality of Experince (QoE) protocols.

\begin{figure}[ht]
    \centering
    \includegraphics[width=0.7\columnwidth]{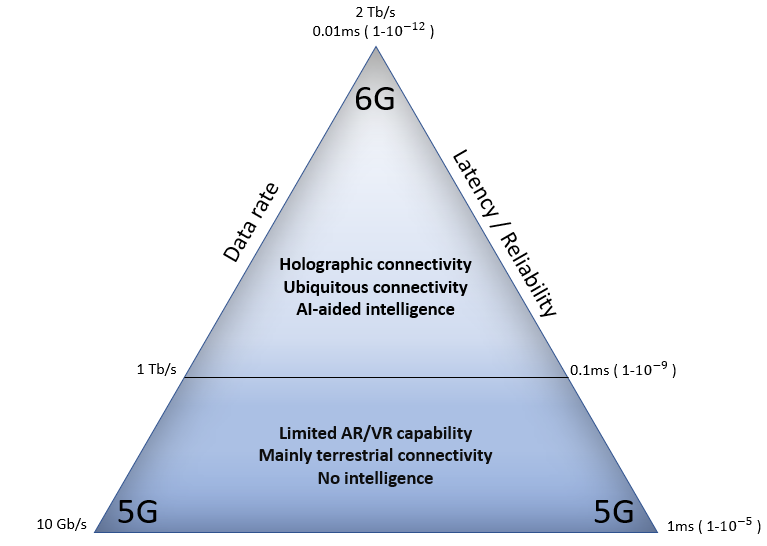}
	\caption{6G Network advantages in data rate and latency compared to 5G networks}
	\label{fig:fig2}
\end{figure}

Cloud-based infrastructure can provide a scalable and flexible solution for VR collaboration networking, using cloud-based servers to handle the processing and rendering of VR content, or using cloud-based services to provide low-latency networking and connectivity \cite{abdelrazeq2020cloud}. Meanwhile, dedicated high-speed networks can provide a reliable and low-latency solution for VR collaboration networking, using dedicated fiber-optic or high-speed wireless networks to provide the required bandwidth and latency for VR collaboration \cite{9488780}.

Edge computing can help to reduce latency in VR collaboration networking by processing data closer to the user's device, using edge computing nodes or devices to handle the processing and rendering of VR content, or using edge-based networking solutions to provide low-latency connectivity \cite{peijoint}. Network optimization techniques can also improve the performance of VR collaboration networking by reducing latency and improving bandwidth utilization, using network optimization tools or techniques to prioritize VR traffic, reduce packet loss, and improve network congestion control \cite{7243304}.

Finally, QoS / QoE protocols can prioritize VR traffic over other types of traffic on the network, ensuring that VR collaboration receives the necessary bandwidth and latency to function effectively. This can involve using QoS protocols such as Differentiated Services (DiffServ) or Resource Reservation Protocol (RSVP) \cite{9603876, iot5020011} to prioritize VR traffic over other types of traffic \cite{krogfoss2020quantifying}.

By employing these solutions for VR collaboration networking, designers and developers can create reliable, low-latency, and high-bandwidth networking solutions that can support effective virtual reality collaboration \cite{10.1117/12.3013532}. These solutions can help to address the technical and practical challenges of VR collaboration, and ensure that users can collaborate effectively and efficiently in virtual reality environments.

\subsection{Automatic Interpupillary Distance Adjustment}

To make HMDs more accessible to the general public, it is crucial to develop a multi-user-customized HMD that can automatically adjust the distance between the left and right eyes to accommodate each user. The binocular distance, which is the distance between the centers of the pupils, can be used to define this distance for each individual \cite{aukstakalnis2016practical}. Nonetheless, the majority of commercially available HMD devices are engineered to accommodate users with an inter-pupillary distance (IPD) of 65mm only. Given that the IPDs of general HMD users vary between 51mm to 77mm, inconsistency in 3D content, which is viewed using both eyes, can result in adverse effects such as dizziness, nausea, vomiting, and eye fatigue. Automatic interpupillary distance (IPD) adjustment \cite{lee2022evaluation} is a feature in some head-mounted display (HMD) devices that allows for the automatic calibration of the distance between the left and right eye display based on the individual user's IPD. This feature helps to ensure that the stereo images viewed with both eyes are correctly aligned, which can prevent adverse effects such as dizziness, nausea, and eye strain. The automatic IPD adjustment feature can enhance the user's overall experience and comfort while using an HMD device.

\subsection{Training Effectiveness} 

Our review found that VR training is as good as, if not better than, traditional in-person training methods. As a result, organizations can use VR training to enhance employee productivity in the workplace, particularly by leveraging its capacity for creating realistic simulations. For instance, instead of using fake phishing emails to educate employees on security, organizations can use VR to instill fear in their Security Education, Training, and Awareness (SETA) programs \cite{dincelli2020choose}. By using interactive metaphors and realistic scenarios that show the consequences of clicking on a phishing email, employees can improve their efficiency and reduce errors by practicing these scenarios multiple times. Additionally, VR enables organizations to tailor their training material for each employee, resulting in a more personalized learning experience in a simulated environment. While tailored training is not unique to VR, the malleability of the virtual environment adds an extra layer of personalization to the training content, which is crucial for improving training effectiveness. Moreover, the virtual environment can be designed to reduce cognitive load or restore attention, which can improve employee responsiveness to the training materials. For example, including nature scenes can improve employee responses to situational factors, such as anxiety, stress, and fear, as well as emergency situations. VR also enables in-depth data collection during training sessions using sensory devices like EEG, eye trackers, and GSR, which can further enhance training programs' personalization.

\subsection{Display}
Displays are an essential component of collaboration in immersive environments. The effectiveness of collaboration in immersive environments depends heavily on the quality and functionality of the displays used to visualize virtual objects and environments \cite{jang2019progress, zhan2020augmented}. There are several methodologies for VR displays that are used to optimize the visual experience and create a more immersive and engaging VR environment.

\begin{figure}[ht]
    \centering
    \includegraphics[width=1\columnwidth]{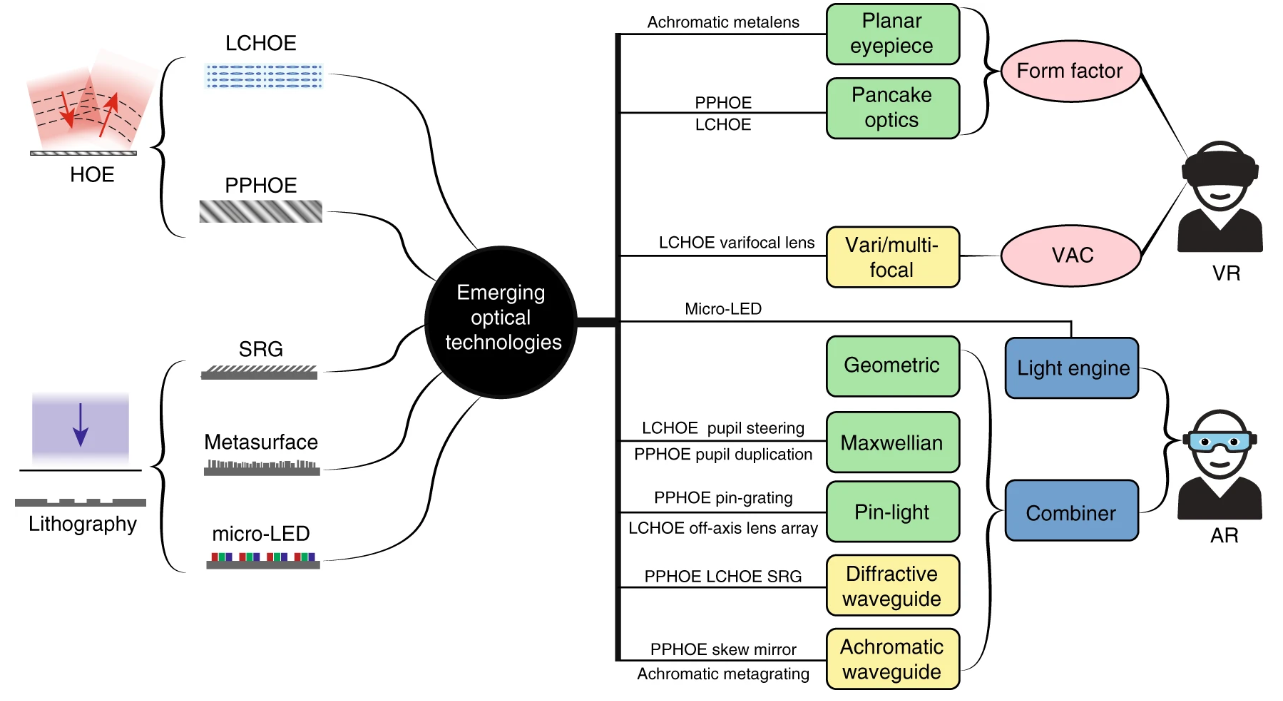}
	\caption{Schematic of some emerging optical technologies applied in AR/VR from \cite{xiong2021augmented}}
	\label{fig:fig4}
\end{figure}

Foveated rendering uses eye-tracking to determine where the user is looking and renders high-quality images only in that area, reducing the amount of processing power needed for VR displays, making it more efficient and enabling higher quality graphics \cite{kim2019foveated}. Distortion correction methodologies can help to correct the distortion that can occur due to the curved lenses used in VR headsets, ensuring that the virtual environment is displayed accurately and reducing the possibility of visual artifacts \cite{li2020distortion}. Asynchronous timewarp uses data from previous frames to create a new frame, reducing latency and improving the overall smoothness of the VR display \cite{hong202256}. Variable rate shading reduces the number of pixels rendered in areas of the VR display that are less important, such as the edges of the display. This methodology can help to improve the efficiency of VR displays, enabling higher quality graphics and smoother performance \cite{iseland2020evaluation}. Display calibration is important to ensure that colors, brightness, and contrast are accurately represented in the VR collaboration display. This may involve using calibration tools or software to adjust display settings for optimal visual fidelity.

Collaborative display environments, such as large-scale projection displays or multi-display systems, can help to create a more engaging and collaborative VR collaboration experience. Projection mapping or other techniques can be used to create a seamless and immersive display environment.

Visual collaboration tools, such as whiteboarding or annotation tools, can help to enhance collaboration and communication among team members in VR collaboration environments \cite{trillmichevolution}. These tools enable users to annotate virtual objects, make notes, and share ideas in real-time, which can improve the overall collaboration experience.

\subsection{Social presence}

Social presence is particularly important for collaboration in virtual reality, as it can help to create a sense of shared presence and enhance communication and collaboration among remote users. To enhance social presence and collaboration in virtual reality, there are several important factors to consider.

Creating realistic avatars can enhance social presence by allowing users to represent themselves in a more natural and expressive way, helping to create a sense of community and shared purpose \cite{9598519}. Additionally, natural and intuitive interactions, such as natural gestures and movements, can help to create a sense of presence and enhance communication \cite{gamelin2021point}.

Collaborating on shared goals and tasks can help to create a sense of community and shared purpose, increasing social presence and enhancing collaboration. By working together towards a common goal, users can feel more connected and engaged with each other \cite{ye2019enhancing}.

High-quality audio and video communication can enhance social presence by allowing users to see and hear each other as they would in the real world. This can help to create a sense of shared presence and enhance communication and collaboration \cite{wen2020photorealistic}. Providing feedback and engagement can enhance social presence by helping users to feel that they are part of a larger community and that their contributions are valued \cite{park2019study}. By providing feedback and engaging with users, designers and developers can create a more engaging and social experience in virtual reality collaboration.

Overall, by focusing on creating realistic avatars, natural interactions, shared goals and tasks, audio and video communication, and providing feedback and engagement, virtual reality collaboration can become a more engaging and social experience for users.

\subsection{Usability}
Navigating and interacting with virtual environments can indeed be difficult, particularly for non-technical users who may not be familiar with the interface and the technology \cite{derby2021challenges}. However, there are several strategies and best practices that can make virtual environment navigation and interaction easier for all users, regardless of their technical expertise. Here are a few examples:

Designers should aim to create navigation interfaces that are intuitive and easy to understand. For example, using familiar and recognizable icons and labels can help users to understand the navigation options available to them \cite{kamarulzaman2020comparative}. Designers should aim to create interactions that are easy to use and understand. For example, using simple and consistent controls, such as buttons or sliders, can make it easier for users to interact with virtual environments.

Providing tutorials and guides can help users to understand how to navigate and interact with virtual environments. For example, including a tutorial or guide at the beginning of the experience can help users to become familiar with the interface and technology \cite{whitlock2020authar}. Providing contextual feedback can help users to understand how to navigate and interact with virtual environments. For example, providing visual \cite{9975360, make6010010} or audio feedback when a button is pressed can help users to understand that an action has been taken.

To create an engaging and inclusive virtual reality experience, designers should focus on creating intuitive and accessible virtual environments that are easy to navigate and interact with. This can involve using familiar icons and labels, providing simple and consistent controls, offering tutorials and guides, providing contextual feedback, and creating accessibility features for users with disabilities. By designing with the needs of all users in mind, designers can create virtual environments that are user-friendly, accessible, and engaging for everyone. Overall, the goal is to create a virtual reality experience that is inclusive and engaging, providing a sense of presence and enhancing collaboration among users.

\subsection{Tracking}

 Tracking refers to the technology used to track the movement and position of the user's head and hands in virtual reality environments. It allows users to move around and interact with the virtual environment in a way that feels natural and intuitive, enhancing the overall experience of VR collaboration. Tracking is a critical component of VR collaboration, enabling users to interact with virtual objects and environments in a natural and intuitive way \cite{4637362}. Here are some of the most common methodologies used for VR tracking:

Designers and developers can make the most of 6DoF tracking in collaborative VR by ensuring that the virtual environment is designed to support natural and intuitive interactions. This can involve using physics engines or other simulation tools to create a more realistic and immersive virtual environment with virtual objects and elements that respond to the user's movements in a realistic way. Additionally, designers and developers should consider how to incorporate collaboration tools and features that take advantage of 6DoF tracking. For example, whiteboarding or annotation tools that allow users to draw or write in virtual space can be a powerful collaboration tool in a 6DoF environment \cite{9146929}. By combining these approaches, designers and developers can create collaborative VR environments that are more natural, intuitive, and engaging for users, enhancing the overall experience of VR collaboration \cite{10.1145/3126594.3126664}.

Inside-out tracking uses cameras or sensors on the VR headset itself to track the user's position and movement, providing a more portable and flexible tracking solution that is less prone to occlusion \cite{eger2020measuring}. Optical tracking uses cameras or sensors to track the position and movement of the user's head and hands, providing a highly accurate and responsive tracking solution that is less prone to interference from other electronic devices \cite{sorriento2019optical}.

Inertial Measurement Unit (IMU) tracking uses sensors to measure the user's movement and acceleration, enabling tracking without the need for external cameras or sensors. This methodology is particularly useful for portable VR systems or applications where external tracking may not be feasible \cite{semwal2022pattern}.

Magnetic tracking uses magnetic sensors to track the user's position and movement, providing a highly accurate and responsive tracking solution that is particularly useful for applications where optical tracking may not be feasible due to environmental factors such as low lighting \cite{9085357}.

By employing these tracking solutions for VR collaboration, designers and developers can create immersive and engaging collaborative environments that enable natural and intuitive interactions. These solutions can help to address the technical and practical challenges of VR collaboration, and ensure that users can collaborate effectively and efficiently in virtual reality and augmented reality environments.

\subsection{Reducing cost}
The cost of collaboration in VR can be a significant barrier for some users. VR technology can be expensive, requiring high-end headsets, powerful computers, and other specialized equipment. This can make it difficult for individuals or small organizations to afford the necessary hardware and software.

In addition to the cost of the equipment, there may also be additional expenses associated with collaborating in VR, such as software licenses or maintenance fees. These costs can add up quickly, making it challenging for some users to justify the expense. Another factor to consider is the cost of training users to use the technology effectively. VR collaboration tools often require a certain level of technical expertise, which can be a barrier for users who are not familiar with the technology.

Because of its affordances for creation, embodiment, and engagement, VR offers several options for businesses to cut operational costs \cite{kodeboyina2016low}. Examples of areas where VR can create opportunities to reduce tangible and intangible costs include the cost of training sessions \cite{farra2019comparative}, costs related to design and maintenance errors due to rework and repairs, and costs that are created by externalities associated with low-quality and defective products, such as reputation and production time. Similar to how it may be utilized as a remote platform to cut down on trip costs for teamwork. It's vital to remember that VR only slightly reduces costs. Particularly the intangible advantages, such as efficient teamwork, higher-quality goods and services, improved customer experiences, and safer training settings for their staff, will eventually be achieved. Even while inexpensive immersive VR systems have the potential to be just as effective as more expensive options, the initial cost of constructing VR systems will almost certainly make the problem of marginal and long-term benefit worse \cite{parham2019creating}. Yet, as VR technology develops, it might eventually be less expensive to maintain a virtual good or service.

In the next section of our discussion, we will be examining and analyzing various examples of how a particular concept, theory, tool, or technology can be used in practical situations. This could involve exploring case studies, real-world examples, or hypothetical scenarios to illustrate the potential uses and benefits of AR/VR Collaboration. The purpose of this section could be to provide a more comprehensive understanding of the topic, demonstrate its relevance and importance, or offer insights into potential applications that may not have been previously considered. By exploring different applications, we can gain a deeper appreciation for the versatility and adaptability of the subject, as well as identify areas for further development or improvement.


\section{Applications}
Collaboration in Immersive environments has become increasingly popular in recent years, and it has numerous applications in various industries. It has become increasingly popular in recent years, particularly with the advancements in augmented reality (AR) and virtual reality (VR) technologies. Immersive environments provide a shared space where people can collaborate, communicate, and interact with each other in real-time, regardless of their physical location. Thus far, we have discussed the challenges and potential solutions related to the use of VR and AR for collaboration in the previous sections. In the following section, we will explore some of the most prevalent applications of VR collaboration:

\begin{figure}[ht]
    \centering
    \includegraphics[width=0.5\columnwidth]{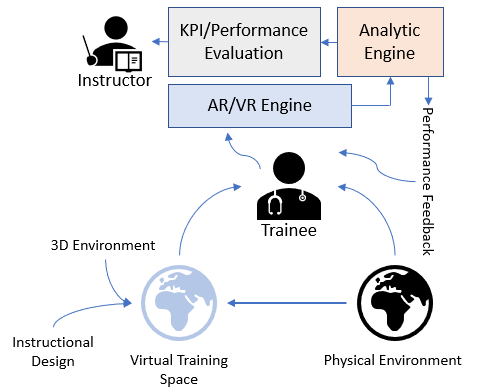}
	\caption{System architecture of an AR/VR training framework}
	\label{fig:fig5}
\end{figure}

\subsection{Training and Education}

AR has enormous potential in training, as it can provide an immersive and engaging experience that helps learners retain information more effectively \cite{sorko2019potentials}. VR collaboration can also be used in training and education to create immersive learning experiences. For example, medical students can practice surgical procedures in a virtual environment before performing them on real patients \cite{8942324}.

Remote learning has also become increasingly popular in recent years, particularly due to the COVID-19 pandemic, which forced many educational institutions to switch to remote learning to ensure the safety of students and teachers. Although remote learning offers several benefits, such as flexibility and accessibility, it also poses some challenges, which were discussed earlier in the challenges section (part 2.5). AR technology can provide learners with a more immersive and engaging learning experience by superimposing digital content onto the real world. This allows learners to interact with virtual objects and environments in real-time, making the learning experience more interactive and memorable \cite{kapp2022design, 9619037}. VR collaboration has the potential to transform training and education by providing immersive and engaging learning experiences for learners. By creating safe and controlled environments for learners to practice their skills, VR collaboration can help to improve learning outcomes and prepare learners for real-world situations.

In addition, AR can be used to overlay digital information, graphics, or animations onto the real-world environment, providing athletes with additional information, feedback, or guidance during training sessions. For example, AR can be used to display virtual markers or targets for athletes to aim at during shooting or throwing exercises, or to provide real-time feedback on the athlete's form or technique \cite{li2021application}. Additionally, AR can be used to simulate game scenarios, allowing athletes to practice decision-making, situational awareness, and teamwork in a realistic environment  \cite{10.1145/3411764.3445649}. AR technology can enhance the effectiveness and efficiency of sports training, enabling athletes to improve their skills and performance in a more engaging and immersive manner.

In conclusion, AR and VR collaboration technologies have enormous potential in training and education, providing learners with engaging and immersive experiences that can improve learning outcomes. By addressing the challenges of remote learning, AR and VR collaboration can revolutionize the way we learn and prepare learners for the future.
 
\subsection{Healthcare}
VR collaboration has the potential to revolutionize the healthcare industry by providing innovative solutions for patient care, medical training, and telemedicine \cite{thomason2021metahealth}. One solution that  has become increasingly popular in recent years is Telemedicine. Telemedicine is the practice of providing medical care and services remotely using technology, which can offer patients a convenient and accessible way to access healthcare from their own homes. For instance, telemedicine can be used to monitor patients with chronic conditions \cite{horrell2021telemedicine}, such as diabetes, by allowing healthcare professionals to remotely monitor blood sugar levels and adjust treatment plans as necessary. This can lead to improved patient outcomes and a better quality of life for patients. Despite the challenges associated with telemedicine, it has the potential to revolutionize the way we access medical care by providing a more convenient and efficient alternative to traditional in-person visits.

\begin{figure}[ht]
    \centering
    \includegraphics[width=1\columnwidth]{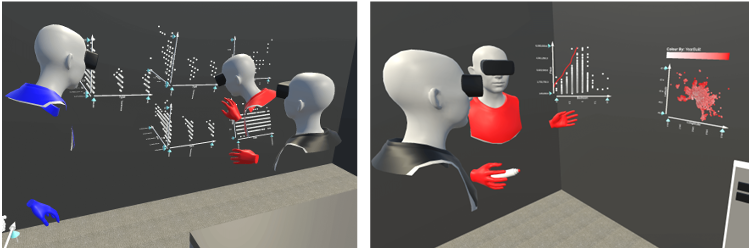}
	\caption{Application of Collaborative Data Visualisation in a Co-located Immersive Environment from \cite{9222346}}
	\label{fig:fig6}
\end{figure}

Rehabilitation is another application of virtual environments to create immersive and engaging exercises for patients recovering from injuries or surgeries. Patients can work with their therapists in a virtual environment, which can help to increase motivation and adherence to treatment plans \cite{pedram2020examining}.

Mental Health: VR collaboration can be used in mental health to provide virtual therapy sessions for patients. Patients can interact with virtual environments and avatars, which can help to reduce anxiety and improve engagement in therapy.

VR collaboration can be used in medical research to create virtual environments for experiments and simulations. Researchers can collaborate in real-time to analyze data and develop new treatments. VR can offer researchers a more immersive and interactive way to explore, analyze and visualize data \cite{eckert2019augmented}. As the technology continues to evolve, we can expect to see even more innovative uses of VR in medical research, leading to new discoveries and breakthroughs in the field.


\subsection{Industry}

VR collaboration has wide-ranging applications for virtual design and prototyping in various industries, providing designers and engineers with a more immersive and interactive way to collaborate and iterate on 3D models in real-time. This can be especially useful in industries such as architecture, engineering, and product design, where the ability to visualize and refine designs in real-time can significantly improve the design process and reduce errors \cite{guo2020applications}. On the other hand, immersive reality has numerous applications in different industries, ranging from architecture and construction to entertainment and gaming, among others. In the architecture and construction industry, immersive reality can provide designers, architects, and builders with a better understanding of the design concepts, allowing them to visualize and explore 3D models and make more informed decisions \cite{8713454}.

Both VR collaboration and immersive reality have numerous applications in sales and marketing, providing companies with more immersive and engaging ways to showcase their products and services. For example, VR collaboration can be used in sales and marketing to visualize and demonstrate products in a more realistic and interactive way. Sales representatives can collaborate with designers and engineers to create virtual prototypes that allow customers to experience products and services in a more immersive and engaging way \cite{qin2021virtual}. This can provide customers with a better understanding of the product, leading to increased sales and customer satisfaction. Moreover, both VR collaboration and immersive reality can provide personalized experiences for customers. In VR collaboration, customers can customize products and experiences to their individual preferences and needs, while immersive reality can create personalized marketing campaigns that target specific demographics and interests.

In the automotive industry, both VR collaboration and immersive reality have various applications, providing designers and engineers with new and innovative ways to design, test, and showcase vehicles. VR collaboration can be used for virtual design and prototyping of vehicles, allowing designers and engineers to collaborate and iterate on 3D models in real-time. This can help streamline the design and prototyping process, reduce errors, and enhance the final product. Additionally, VR collaboration can be used for remote collaboration, allowing designers and engineers from different locations to work together on the same project.

\subsection{Entertainment and Gaming}

Entertainment and gaming are the industries where virtual reality (VR) and augmented reality (AR) technologies first gained significant traction \cite{bates1992virtual}. VR and AR can provide users with a more immersive and interactive gaming experience, allowing them to enter virtual worlds and interact with digital content in a more realistic and engaging way.

In gaming, VR can provide users with a more immersive and realistic gaming experience, allowing them to explore virtual worlds and interact with digital content in a more realistic way. VR can also provide users with a greater sense of presence and agency within the game, making the experience more engaging and enjoyable \cite{pallavicini2019gaming}.

AR, on the other hand, can provide users with a more interactive and engaging gaming experience, allowing them to overlay digital content onto the real world. This can create unique and innovative gaming experiences, such as augmented reality games that use the player's physical environment as part of the game \cite{shin2019does}.

Moreover, VR and AR can also provide new and innovative ways to interact with entertainment content, such as movies and television shows. For example, VR can provide users with an immersive and interactive movie experience, allowing them to enter the movie and interact with the characters and environment. AR, on the other hand, can be used to overlay digital content onto live performances or events, creating a more engaging and interactive experience for audiences \cite{9223669}.

Overall, VR has transformed the entertainment and gaming industries by providing users with a more immersive and interactive experience. As the technology continues to evolve, we can expect to see even more innovative uses of VR in entertainment and gaming, leading to new and exciting experiences for user


\section{Conclusion}

Collaboration in immersive environments has the potential to revolutionize the way people work and interact with each other. By allowing users to interact with each other and their environment in a more natural and intuitive way, immersive environments can increase productivity, creativity, and collaboration.

Despite the challenges of navigating and interacting with virtual environments, as well as the cost associated with accessing VR technology, the benefits of collaboration in immersive environments are clear. It allows individuals and teams to work together from anywhere in the world, breaking down geographical barriers and enabling more diverse and inclusive teams.

As VR technology continues to advance and become more accessible, we can expect to see even more innovative ways of collaborating in immersive environments. From virtual meeting spaces to collaborative design tools, immersive environments have the potential to transform the way we work and collaborate for years to come.

Collaboration in immersive environments has the potential to transform the way people work, learn, and interact with each other. By combining the power of virtual reality, augmented reality, and other emerging technologies, immersive environments enable users to interact with each other and their surroundings in a more natural and intuitive way.

One of the main advantages of collaboration in immersive environments is that it enables individuals and teams to work together from anywhere in the world. This is particularly important in today's globalized and increasingly remote workforce. Immersive environments can break down geographical barriers and enable more diverse and inclusive teams, which can ultimately lead to more innovative and effective solutions.

Immersive environments also offer a wide range of tools and features that can enhance collaboration and communication. For example, virtual whiteboards, shared work spaces, and collaborative design tools can help teams to brainstorm and iterate on ideas in real-time, regardless of their location. These tools can also be particularly useful for remote teams that may not have access to the same physical resources and equipment.

However, there are also several challenges associated with collaboration in immersive environments. One of the main challenges is the technical complexity of the technology. Immersive environments require specialized hardware and software, which can be difficult and expensive to set up and maintain. Additionally, users may need to undergo extensive training in order to use the technology effectively. Another challenge is the potential for social isolation and disconnection. Immersive environments can be isolating and can make users feel disconnected from the physical world. This can be particularly problematic for individuals who already struggle with feelings of loneliness or social anxiety.

Despite these challenges, the benefits of collaboration in immersive environments are clear. Immersive environments can increase productivity, creativity, and collaboration, while also breaking down geographical barriers and enabling more diverse and inclusive teams. As VR technology continues to advance and become more accessible, we can expect to see even more innovative ways of collaborating in immersive environments. From virtual meeting spaces to collaborative design tools, immersive environments have the potential to transform the way we work and collaborate for years to come.

A leader/facilitator is crucial in immersive collaborative work environments for communication, coordination, and decision-making. They can set goals, establish rules, manage conflicts, and provide guidance to ensure a common goal. A supportive, approachable leader/facilitator fosters a positive and productive work environment, while a controlling one undermines collaboration. The role of the leader/facilitator varies based on the specific immersive environment, requiring adjustments to accommodate technology or physical space.
Overall, while the role of leader/facilitator presence and behavior on collaborative work outcomes in immersive environments is important, there is still a need for more research in this area. As immersive technology continues to evolve, it is important to explore the unique challenges and opportunities it presents for collaborative work and the role of leaders/facilitators in these environments. With further research, we can gain a deeper understanding of how to optimize leadership and collaboration in immersive environments and improve outcomes for teams working in these settings.










\bibliographystyle{IEEEtran}
\bibliography{References}

\begin{thebibliography}{100}
\providecommand{\url}[1]{#1}
\csname url@samestyle\endcsname
\providecommand{\newblock}{\relax}
\providecommand{\bibinfo}[2]{#2}
\providecommand{\BIBentrySTDinterwordspacing}{\spaceskip=0pt\relax}
\providecommand{\BIBentryALTinterwordstretchfactor}{4}
\providecommand{\BIBentryALTinterwordspacing}{\spaceskip=\fontdimen2\font plus
\BIBentryALTinterwordstretchfactor\fontdimen3\font minus
  \fontdimen4\font\relax}
\providecommand{\BIBforeignlanguage}[2]{{%
\expandafter\ifx\csname l@#1\endcsname\relax
\typeout{** WARNING: IEEEtran.bst: No hyphenation pattern has been}%
\typeout{** loaded for the language `#1'. Using the pattern for}%
\typeout{** the default language instead.}%
\else
\language=\csname l@#1\endcsname
\fi
#2}}
\providecommand{\BIBdecl}{\relax}
\BIBdecl

\bibitem{10.1145/77276.77278}
\BIBentryALTinterwordspacing
C.~Conn, J.~Lanier, M.~Minsky, S.~Fisher, and A.~Druin, ``Virtual environments
  and interactivity: Windows to the future,'' in \emph{ACM SIGGRAPH 89 Panel
  Proceedings}, ser. SIGGRAPH '89.\hskip 1em plus 0.5em minus 0.4em\relax New
  York, NY, USA: Association for Computing Machinery, 1989, p. 7–18.
  [Online]. Available: \url{https://doi.org/10.1145/77276.77278}
\BIBentrySTDinterwordspacing

\bibitem{250922}
J.~Encarnacao, M.~Gobel, and L.~Rosenblum, ``European activities in virtual
  reality,'' \emph{IEEE Computer Graphics and Applications}, vol.~14, no.~1,
  pp. 66--74, 1994.

\bibitem{alaker2016virtual}
M.~Alaker, G.~R. Wynn, and T.~Arulampalam, ``Virtual reality training in
  laparoscopic surgery: a systematic review \& meta-analysis,''
  \emph{International Journal of Surgery}, vol.~29, pp. 85--94, 2016.

\bibitem{9757602}
S.~Doroudian, Z.~Wu, W.~Wang, A.~Galati, and A.~Lu, ``A study of real-time
  information on user behaviors during search and rescue (sar) training of
  firefighters,'' in \emph{2022 IEEE Conference on Virtual Reality and 3D User
  Interfaces Abstracts and Workshops (VRW)}, 2022, pp. 387--394.

\bibitem{8797889}
R.~M. Clifford, S.~Jung, S.~Hoermann, M.~Billinghurst, and R.~W. Lindeman,
  ``Creating a stressful decision making environment for aerial firefighter
  training in virtual reality,'' in \emph{2019 IEEE Conference on Virtual
  Reality and 3D User Interfaces (VR)}, 2019, pp. 181--189.

\bibitem{10.2307/2185537}
\BIBentryALTinterwordspacing
M.~E. Bratman, ``Shared cooperative activity,'' \emph{The Philosophical
  Review}, vol. 101, no.~2, pp. 327--341, 1992. [Online]. Available:
  \url{http://www.jstor.org/stable/2185537}
\BIBentrySTDinterwordspacing

\bibitem{1352977}
V.~Theoktisto and M.~Fairen, ``On extending collaboration in virtual reality
  environments,'' in \emph{Proceedings. 17th Brazilian Symposium on Computer
  Graphics and Image Processing}, 2004, pp. 324--331.

\bibitem{9757558}
A.~Bayro, Y.~Ghasemi, and H.~Jeong, ``Subjective and objective analyses of
  collaboration and co-presence in a virtual reality remote environment,'' in
  \emph{2022 IEEE Conference on Virtual Reality and 3D User Interfaces
  Abstracts and Workshops (VRW)}, 2022, pp. 485--487.

\bibitem{10.1162/pres.1997.6.6.603}
\BIBentryALTinterwordspacing
M.~Slater and S.~Wilbur, ``A framework for immersive virtual environments five:
  Speculations on the role of presence in virtual environments,''
  \emph{Presence: Teleoper. Virtual Environ.}, vol.~6, no.~6, p. 603–616, dec
  1997. [Online]. Available: \url{https://doi.org/10.1162/pres.1997.6.6.603}
\BIBentrySTDinterwordspacing

\bibitem{10.1145/3491102.3517500}
\BIBentryALTinterwordspacing
N.~Cila, ``Designing human-agent collaborations: Commitment, responsiveness,
  and support,'' in \emph{Proceedings of the 2022 CHI Conference on Human
  Factors in Computing Systems}, ser. CHI '22.\hskip 1em plus 0.5em minus
  0.4em\relax New York, NY, USA: Association for Computing Machinery, 2022.
  [Online]. Available: \url{https://doi.org/10.1145/3491102.3517500}
\BIBentrySTDinterwordspacing

\bibitem{10.1145/1841853.1841856}
\BIBentryALTinterwordspacing
S.~Lewis, J.~B. Ellis, and W.~A. Kellogg, ``Using virtual interactions to
  explore leadership and collaboration in globally distributed teams,'' in
  \emph{Proceedings of the 3rd International Conference on Intercultural
  Collaboration}, ser. ICIC '10.\hskip 1em plus 0.5em minus 0.4em\relax New
  York, NY, USA: Association for Computing Machinery, 2010, p. 9–18.
  [Online]. Available: \url{https://doi.org/10.1145/1841853.1841856}
\BIBentrySTDinterwordspacing

\bibitem{park2019literature}
M.~J. Park, D.~J. Kim, U.~Lee, E.~J. Na, and H.~J. Jeon, ``A literature
  overview of virtual reality (vr) in treatment of psychiatric disorders:
  recent advances and limitations,'' \emph{Frontiers in psychiatry}, vol.~10,
  p. 505, 2019.

\bibitem{chheang2021collaborative}
V.~Chheang, P.~Saalfeld, F.~Joeres, C.~Boedecker, T.~Huber, F.~Huettl, H.~Lang,
  B.~Preim, and C.~Hansen, ``A collaborative virtual reality environment for
  liver surgery planning,'' \emph{Computers \& Graphics}, vol.~99, pp.
  234--246, 2021.

\bibitem{fu2022systematic}
Y.~Fu, Y.~Hu, and V.~Sundstedt, ``A systematic literature review of virtual,
  augmented, and mixed reality game applications in healthcare,'' \emph{ACM
  Transactions on Computing for Healthcare (HEALTH)}, vol.~3, no.~2, pp. 1--27,
  2022.

\bibitem{6225733}
C.~Chen, S.~Helal, S.~de~Deugd, A.~Smith, and C.~K. Chang, ``Toward a
  collaboration model for smart spaces,'' in \emph{2012 Third International
  Workshop on Software Engineering for Sensor Network Applications (SESENA)},
  2012, pp. 37--42.

\bibitem{8009379}
L.~Devigne, M.~Babel, F.~Nouviale, V.~K. Narayanan, F.~Pasteau, and P.~Gallien,
  ``Design of an immersive simulator for assisted power wheelchair driving,''
  in \emph{2017 International Conference on Rehabilitation Robotics (ICORR)},
  2017, pp. 995--1000.

\bibitem{9982707}
E.~Cibuļska and K.~Boločko, ``Virtual reality in education: Structural design
  of an adaptable virtual reality system,'' in \emph{2022 6th International
  Conference on Computer, Software and Modeling (ICCSM)}, 2022, pp. 76--79.

\bibitem{9524249}
H.~Liu, ``Research on the development of 5g industry assisted by virtual
  reality technology,'' in \emph{2021 IEEE International Conference on Power,
  Intelligent Computing and Systems (ICPICS)}, 2021, pp. 480--483.

\bibitem{gerup2020augmented}
J.~Gerup, C.~B. Soerensen, and P.~Dieckmann, ``Augmented reality and mixed
  reality for healthcare education beyond surgery: an integrative review,''
  \emph{International journal of medical education}, vol.~11, p.~1, 2020.

\bibitem{hung2021new}
S.-W. Hung, C.-W. Chang, and Y.-C. Ma, ``A new reality: Exploring continuance
  intention to use mobile augmented reality for entertainment purposes,''
  \emph{Technology in Society}, vol.~67, p. 101757, 2021.

\bibitem{10.1145/3313831.3376722}
\BIBentryALTinterwordspacing
N.~Ashtari, A.~Bunt, J.~McGrenere, M.~Nebeling, and P.~K. Chilana, ``Creating
  augmented and virtual reality applications: Current practices, challenges,
  and opportunities,'' in \emph{Proceedings of the 2020 CHI Conference on Human
  Factors in Computing Systems}, ser. CHI '20.\hskip 1em plus 0.5em minus
  0.4em\relax New York, NY, USA: Association for Computing Machinery, 2020, p.
  1–13. [Online]. Available: \url{https://doi.org/10.1145/3313831.3376722}
\BIBentrySTDinterwordspacing

\bibitem{digirolamo2019exploration}
J.~A. Digirolamo and J.~T. Tkach, ``An exploration of managers and leaders
  using coaching skills.'' \emph{Consulting Psychology Journal: Practice and
  Research}, vol.~71, no.~3, p. 195, 2019.

\bibitem{jenifer2015cross}
R.~D. Jenifer and G.~Raman, ``Cross-cultural communication barriers in the
  workplace,'' \emph{International Journal of Management}, vol.~6, no.~1, pp.
  348--351, 2015.

\bibitem{huang2021interactivity}
Y.~Huang, D.~Gursoy, M.~Zhang, R.~Nunkoo, and S.~Shi, ``Interactivity in online
  chat: Conversational cues and visual cues in the service recovery process,''
  \emph{International Journal of Information Management}, vol.~60, p. 102360,
  2021.

\bibitem{tannen1983cross}
D.~Tannen, ``Cross-cultural communication.'' \emph{Corporate Communications: an
  international journal}, 1983.

\bibitem{colleoni2013csr}
E.~Colleoni, ``Csr communication strategies for organizational legitimacy in
  social media,'' \emph{Corporate Communications: an international journal},
  vol.~18, no.~2, pp. 228--248, 2013.

\bibitem{van2010survey}
D.~Van~Krevelen and R.~Poelman, ``A survey of augmented reality technologies,
  applications and limitations,'' \emph{International journal of virtual
  reality}, vol.~9, no.~2, pp. 1--20, 2010.

\bibitem{8743304}
M.~Fiedler, H.-J. Zepernick, and V.~Kelkkanen, ``Network-induced temporal
  disturbances in virtual reality applications,'' in \emph{2019 Eleventh
  International Conference on Quality of Multimedia Experience (QoMEX)}, 2019,
  pp. 1--3.

\bibitem{pellas2019augmenting}
N.~Pellas, P.~Fotaris, I.~Kazanidis, and D.~Wells, ``Augmenting the learning
  experience in primary and secondary school education: A systematic review of
  recent trends in augmented reality game-based learning,'' \emph{Virtual
  Reality}, vol.~23, no.~4, pp. 329--346, 2019.

\bibitem{9416950}
IEEE, ``Ieee standard for head-mounted display (hmd)-based virtual reality(vr)
  sickness reduction technology,'' \emph{IEEE Std 3079-2020}, pp. 1--74, 2021.

\bibitem{9757480}
T.~Hopkins, S.~C.-C. Weng, R.~Vanukuru, E.~Wenzel, A.~Banic, and E.~Y.-L. Do,
  ``How late is too late? effects of network latency on audio-visual perception
  during ar remote musical collaboration,'' in \emph{2022 IEEE Conference on
  Virtual Reality and 3D User Interfaces Abstracts and Workshops (VRW)}, 2022,
  pp. 686--687.

\bibitem{9913710}
T.~Dang, C.~Liu, and M.~Peng, ``Low-latency mobile virtual reality content
  delivery for unmanned aerial vehicle-enabled wireless networks with energy
  constraints,'' \emph{IEEE Transactions on Vehicular Technology}, vol.~72,
  no.~2, pp. 2189--2201, 2023.

\bibitem{stauffert2020latency}
J.-P. Stauffert, F.~Niebling, and M.~E. Latoschik, ``Latency and cybersickness:
  Impact, causes, and measures. a review,'' \emph{Frontiers in Virtual
  Reality}, vol.~1, p. 582204, 2020.

\bibitem{9268953}
Y.~Zhou, C.~Pan, P.~L. Yeoh, K.~Wang, M.~Elkashlan, B.~Vucetic, and Y.~Li,
  ``Communication-and-computing latency minimization for uav-enabled virtual
  reality delivery systems,'' \emph{IEEE Transactions on Communications},
  vol.~69, no.~3, pp. 1723--1735, 2021.

\bibitem{10.1145/3485279.3485283}
\BIBentryALTinterwordspacing
D.~Schneider, V.~Biener, A.~Otte, T.~Gesslein, P.~Gagel, C.~Campos,
  K.~\v{C}opi\v{c} Pucihar, M.~Kljun, E.~Ofek, M.~Pahud, P.~O. Kristensson, and
  J.~Grubert, ``Accuracy evaluation of touch tasks in commodity virtual and
  augmented reality head-mounted displays,'' in \emph{Proceedings of the 2021
  ACM Symposium on Spatial User Interaction}, ser. SUI '21.\hskip 1em plus
  0.5em minus 0.4em\relax New York, NY, USA: Association for Computing
  Machinery, 2021. [Online]. Available:
  \url{https://doi.org/10.1145/3485279.3485283}
\BIBentrySTDinterwordspacing

\bibitem{pastel2021comparison}
S.~Pastel, C.-H. Chen, L.~Martin, M.~Naujoks, K.~Petri, and K.~Witte,
  ``Comparison of gaze accuracy and precision in real-world and virtual
  reality,'' \emph{Virtual Reality}, vol.~25, pp. 175--189, 2021.

\bibitem{livitcuka2023impact}
R.~Livitcuka, R.~Alksnis, and T.~Pladere, ``Impact of interpupillary distance
  mismatch on visual aftereffects of virtual reality gameplay,'' in
  \emph{Optical Architectures for Displays and Sensing in Augmented, Virtual,
  and Mixed Reality (AR, VR, MR) IV}, vol. 12449.\hskip 1em plus 0.5em minus
  0.4em\relax SPIE, 2023, pp. 377--380.

\bibitem{9376369}
P.~B. Hibbard, L.~C. van Dam, and P.~Scarfe, ``The implications of
  interpupillary distance variability for virtual reality,'' in \emph{2020
  International Conference on 3D Immersion (IC3D)}, 2020, pp. 1--7.

\bibitem{8797975}
A.~U. Batmaz, M.~D.~B. Machuca, D.~M. Pham, and W.~Stuerzlinger, ``Do
  head-mounted display stereo deficiencies affect 3d pointing tasks in ar and
  vr?'' in \emph{2019 IEEE Conference on Virtual Reality and 3D User Interfaces
  (VR)}, 2019, pp. 585--592.

\bibitem{feng2022wayfinding}
Y.~Feng, D.~C. Duives, and S.~P. Hoogendoorn, ``Wayfinding behaviour in a
  multi-level building: A comparative study of hmd vr and desktop vr,''
  \emph{Advanced Engineering Informatics}, vol.~51, p. 101475, 2022.

\bibitem{ferraz2021benchmarking}
O.~Ferraz, P.~Menezes, V.~Silva, and G.~Falcao, ``Benchmarking vulkan vs opengl
  rendering on low-power edge gpus,'' in \emph{2021 International Conference on
  Graphics and Interaction (ICGI)}.\hskip 1em plus 0.5em minus 0.4em\relax
  IEEE, 2021, pp. 1--8.

\bibitem{gattullo2020informing}
M.~Gattullo, G.~W. Scurati, A.~Evangelista, F.~Ferrise, M.~Fiorentino, and
  A.~E. Uva, ``Informing the use of visual assets in industrial augmented
  reality,'' in \emph{Design Tools and Methods in Industrial Engineering:
  Proceedings of the International Conference on Design Tools and Methods in
  Industrial Engineering, ADM 2019, September 9--10, 2019, Modena,
  Italy}.\hskip 1em plus 0.5em minus 0.4em\relax Springer, 2020, pp. 106--117.

\bibitem{delgado2020research}
J.~M.~D. Delgado, L.~Oyedele, P.~Demian, and T.~Beach, ``A research agenda for
  augmented and virtual reality in architecture, engineering and
  construction,'' \emph{Advanced Engineering Informatics}, vol.~45, p. 101122,
  2020.

\bibitem{de2020survey}
L.~F. de~Souza~Cardoso, F.~C. M.~Q. Mariano, and E.~R. Zorzal, ``A survey of
  industrial augmented reality,'' \emph{Computers \& Industrial Engineering},
  vol. 139, p. 106159, 2020.

\bibitem{10.1145/2543581.2543590}
\BIBentryALTinterwordspacing
R.~S. Renner, B.~M. Velichkovsky, and J.~R. Helmert, ``The perception of
  egocentric distances in virtual environments - a review,'' \emph{ACM Comput.
  Surv.}, vol.~46, no.~2, dec 2013. [Online]. Available:
  \url{https://doi.org/10.1145/2543581.2543590}
\BIBentrySTDinterwordspacing

\bibitem{masnadi2022effects}
S.~Masnadi, K.~Pfeil, J.-V.~T. Sera-Josef, and J.~LaViola, ``Effects of field
  of view on egocentric distance perception in virtual reality,'' in
  \emph{Proceedings of the 2022 CHI Conference on Human Factors in Computing
  Systems}, 2022, pp. 1--10.

\bibitem{trepkowski2019effect}
C.~Trepkowski, D.~Eibich, J.~Maiero, A.~Marquardt, E.~Kruijff, and S.~Feiner,
  ``The effect of narrow field of view and information density on visual search
  performance in augmented reality,'' in \emph{2019 IEEE Conference on Virtual
  Reality and 3D User Interfaces (VR)}.\hskip 1em plus 0.5em minus 0.4em\relax
  IEEE, 2019, pp. 575--584.

\bibitem{wibowo2021improving}
S.~Wibowo, I.~Siradjuddin, F.~Ronilaya, and M.~Hidayat, ``Improving
  teleoperation robots performance by eliminating view limit using 360 camera
  and enhancing the immersive experience utilizing vr headset,'' in \emph{IOP
  Conference Series: Materials Science and Engineering}, vol. 1073.\hskip 1em
  plus 0.5em minus 0.4em\relax IOP Publishing, 2021, p. 012037.

\bibitem{farra2019comparative}
S.~L. Farra, M.~Gneuhs, E.~Hodgson, B.~Kawosa, E.~T. Miller, A.~Simon, N.~Timm,
  and J.~Hausfeld, ``Comparative cost of virtual reality training and live
  exercises for training hospital workers for evacuation,'' \emph{Computers,
  informatics, nursing: CIN}, vol.~37, no.~9, p. 446, 2019.

\bibitem{dincelli2022immersive}
E.~Dincelli and A.~Yayla, ``Immersive virtual reality in the age of the
  metaverse: A hybrid-narrative review based on the technology affordance
  perspective,'' \emph{The journal of strategic information systems}, vol.~31,
  no.~2, p. 101717, 2022.

\bibitem{dris2019openbim}
A.-S. Dris, F.~Lehericey, V.~Gouranton, and B.~Arnaldi, ``Openbim based ive
  ontology: an ontological approach to improve interoperability for virtual
  reality applications,'' in \emph{Advances in Informatics and Computing in
  Civil and Construction Engineering: Proceedings of the 35th CIB W78 2018
  Conference: IT in Design, Construction, and Management}.\hskip 1em plus 0.5em
  minus 0.4em\relax Springer, 2019, pp. 129--136.

\bibitem{shirowzhan2020bim}
S.~Shirowzhan, S.~M. Sepasgozar, D.~J. Edwards, H.~Li, and C.~Wang, ``Bim
  compatibility and its differentiation with interoperability challenges as an
  innovation factor,'' \emph{Automation in Construction}, vol. 112, p. 103086,
  2020.

\bibitem{ohyama2007autonomic}
S.~Ohyama, S.~Nishiike, H.~Watanabe, K.~Matsuoka, H.~Akizuki, N.~Takeda, and
  T.~Harada, ``Autonomic responses during motion sickness induced by virtual
  reality,'' \emph{Auris Nasus Larynx}, vol.~34, no.~3, pp. 303--306, 2007.

\bibitem{10.1145/3411764.3445701}
\BIBentryALTinterwordspacing
C.~MacArthur, A.~Grinberg, D.~Harley, and M.~Hancock, ``You’re making me
  sick: A systematic review of how virtual reality research considers gender ;
  cybersickness,'' in \emph{Proceedings of the 2021 CHI Conference on Human
  Factors in Computing Systems}, ser. CHI '21.\hskip 1em plus 0.5em minus
  0.4em\relax New York, NY, USA: Association for Computing Machinery, 2021.
  [Online]. Available: \url{https://doi.org/10.1145/3411764.3445701}
\BIBentrySTDinterwordspacing

\bibitem{CHEN201517}
\BIBentryALTinterwordspacing
W.~Chen, J.-G. Chao, X.-W. Chen, J.-K. Wang, and C.~Tan, ``Quantitative
  orientation preference and susceptibility to space motion sickness simulated
  in a virtual reality environment,'' \emph{Brain Research Bulletin}, vol. 113,
  pp. 17--26, 2015. [Online]. Available:
  \url{https://www.sciencedirect.com/science/article/pii/S0361923015000246}
\BIBentrySTDinterwordspacing

\bibitem{6496840}
L.~Losik, ``Using the brain's fight-or-flight response for predicting mental
  illness on the human space flight program,'' in \emph{2013 IEEE Aerospace
  Conference}, 2013, pp. 1--20.

\bibitem{lee2021kinect}
I.-J. Lee, ``Kinect-for-windows with augmented reality in an interactive
  roleplay system for children with an autism spectrum disorder,''
  \emph{Interactive Learning Environments}, vol.~29, no.~4, pp. 688--704, 2021.

\bibitem{balbin2019augmented}
J.~R. Balbin, C.~C. Paglinawan, M.~J.~A. de~Castro, J.~K.~C. Llamas, M.~E.~T.
  Medina, J.~J.~O. Pangilinan, and F.~L. Valiente, ``Augmented reality aided
  analysis of customer satisfaction based on taste-induced facial expression
  recognition using affdex software developer's kit,'' in \emph{Proceedings of
  the 2019 9th International Conference on Biomedical Engineering and
  Technology}, 2019, pp. 204--209.

\bibitem{nijholt2022capturing}
A.~Nijholt, ``Capturing obstructed nonverbal cues in augmented reality
  interactions: a short survey,'' in \emph{Proceedings of International
  Conference on Industrial Instrumentation and Control: ICI2C 2021}.\hskip 1em
  plus 0.5em minus 0.4em\relax Springer, 2022, pp. 1--9.

\bibitem{voit2019online}
A.~Voit, S.~Mayer, V.~Schwind, and N.~Henze, ``Online, vr, ar, lab, and
  in-situ: comparison of research methods to evaluate smart artifacts,'' in
  \emph{Proceedings of the 2019 chi conference on human factors in computing
  systems}, 2019, pp. 1--12.

\bibitem{diersch2019potential}
N.~Diersch and T.~Wolbers, ``The potential of virtual reality for spatial
  navigation research across the adult lifespan,'' \emph{Journal of
  Experimental Biology}, vol. 222, no. Suppl\_1, p. jeb187252, 2019.

\bibitem{hejtmanek2020much}
L.~Hejtmanek, M.~Starrett, E.~Ferrer, and A.~D. Ekstrom, ``How much of what we
  learn in virtual reality transfers to real-world navigation?''
  \emph{Multisensory Research}, vol.~33, no. 4-5, pp. 479--503, 2020.

\bibitem{10.1145/3343036.3343131}
\BIBentryALTinterwordspacing
R.~Paris, J.~Klag, P.~Rajan, L.~Buck, T.~P. McNamara, and B.~Bodenheimer, ``How
  video game locomotion methods affect navigation in virtual environments,'' in
  \emph{ACM Symposium on Applied Perception 2019}, ser. SAP '19.\hskip 1em plus
  0.5em minus 0.4em\relax New York, NY, USA: Association for Computing
  Machinery, 2019. [Online]. Available:
  \url{https://doi.org/10.1145/3343036.3343131}
\BIBentrySTDinterwordspacing

\bibitem{yang2019gesture}
L.~Yang, J.~Huang, T.~Feng, W.~Hong-An, and D.~Guo-Zhong, ``Gesture interaction
  in virtual reality,'' \emph{Virtual Reality \& Intelligent Hardware}, vol.~1,
  no.~1, pp. 84--112, 2019.

\bibitem{10.1145/3359626}
\BIBentryALTinterwordspacing
J.~A. De~Guzman, K.~Thilakarathna, and A.~Seneviratne, ``Security and privacy
  approaches in mixed reality: A literature survey,'' \emph{ACM Comput. Surv.},
  vol.~52, no.~6, oct 2019. [Online]. Available:
  \url{https://doi.org/10.1145/3359626}
\BIBentrySTDinterwordspacing

\bibitem{jozani2020privacy}
M.~Jozani, E.~Ayaburi, M.~Ko, and K.-K.~R. Choo, ``Privacy concerns and
  benefits of engagement with social media-enabled apps: A privacy calculus
  perspective,'' \emph{Computers in Human Behavior}, vol. 107, p. 106260, 2020.

\bibitem{ali2018privacy}
S.~Ali, N.~Islam, A.~Rauf, I.~U. Din, M.~Guizani, and J.~J. Rodrigues,
  ``Privacy and security issues in online social networks,'' \emph{Future
  Internet}, vol.~10, no.~12, p. 114, 2018.

\bibitem{harding2019understanding}
E.~L. Harding, J.~J. Vanto, R.~Clark, L.~Hannah~Ji, and S.~C. Ainsworth,
  ``Understanding the scope and impact of the california consumer privacy act
  of 2018,'' \emph{Journal of Data Protection \& Privacy}, vol.~2, no.~3, pp.
  234--253, 2019.

\bibitem{10070393}
A.~M. Aslam, R.~Chaudhary, A.~Bhardwaj, I.~Budhiraja, N.~Kumar, and
  S.~Zeadally, ``Metaverse for 6g and beyond: The next revolution and
  deployment challenges,'' \emph{IEEE Internet of Things Magazine}, vol.~6,
  no.~1, pp. 32--39, 2023.

\bibitem{9656726}
P.~Bhattacharya, D.~Saraswat, A.~Dave, M.~Acharya, S.~Tanwar, G.~Sharma, and
  I.~E. Davidson, ``Coalition of 6g and blockchain in ar/vr space: Challenges
  and future directions,'' \emph{IEEE Access}, vol.~9, pp. 168\,455--168\,484,
  2021.

\bibitem{rashvand2024realtimebusarrivalprediction}
\BIBentryALTinterwordspacing
N.~Rashvand, S.~S. Hosseini, M.~Azarbayjani, and H.~Tabkhi, ``Real-time bus
  arrival prediction: A deep learning approach for enhanced urban mobility,''
  2024. [Online]. Available: \url{https://arxiv.org/abs/2303.15495}
\BIBentrySTDinterwordspacing

\bibitem{giordani2020toward}
M.~Giordani, M.~Polese, M.~Mezzavilla, S.~Rangan, and M.~Zorzi, ``Toward 6g
  networks: Use cases and technologies,'' \emph{IEEE Communications Magazine},
  vol.~58, no.~3, pp. 55--61, 2020.

\bibitem{abdelrazeq2020cloud}
A.~Abdelrazeq, C.~Kohlschein, and F.~Hees, ``A cloud based augmented reality
  framework-enabling user-centered interactive systems development,'' in
  \emph{Human Systems Engineering and Design II: Proceedings of the 2nd
  International Conference on Human Systems Engineering and Design (IHSED2019):
  Future Trends and Applications, September 16-18, 2019, Universit{\"a}t der
  Bundeswehr M{\"u}nchen, Munich, Germany}.\hskip 1em plus 0.5em minus
  0.4em\relax Springer, 2020, pp. 417--422.

\bibitem{9488780}
G.~Naik and J.-M.~J. Park, ``Coexistence of wi-fi 6e and 5g nr-u: Can we do
  better in the 6 ghz bands?'' in \emph{IEEE INFOCOM 2021 - IEEE Conference on
  Computer Communications}, 2021, pp. 1--10.

\bibitem{peijoint}
Y.~Pei, M.~Li, H.~Wu, Q.~Ye, C.~Zhou, S.~Hu, and X.~S. Shen, ``Joint caching
  and computing resource reservation for edge-assisted location-aware augmented
  reality,'' \emph{Arxive}, 2021.

\bibitem{7243304}
R.~Mijumbi, J.~Serrat, J.-L. Gorricho, N.~Bouten, F.~De~Turck, and R.~Boutaba,
  ``Network function virtualization: State-of-the-art and research
  challenges,'' \emph{IEEE Communications Surveys and Tutorials}, vol.~18,
  no.~1, pp. 236--262, 2016.

\bibitem{9603876}
J.~Liu, ``Design and implementation of vo ipqos model combining intserv and
  diffserv based on network processor ixp2400,'' in \emph{2021 7th Annual
  International Conference on Network and Information Systems for Computers
  (ICNISC)}, 2021, pp. 60--64.

\bibitem{iot5020011}
\BIBentryALTinterwordspacing
N.~Rashvand, K.~Witham, G.~Maldonado, V.~Katariya, N.~Marer~Prabhu,
  G.~Schirner, and H.~Tabkhi, ``Enhancing automatic modulation recognition for
  iot applications using transformers,'' \emph{IoT}, vol.~5, no.~2, pp.
  212--226, 2024. [Online]. Available:
  \url{https://www.mdpi.com/2624-831X/5/2/11}
\BIBentrySTDinterwordspacing

\bibitem{krogfoss2020quantifying}
B.~Krogfoss, J.~Duran, P.~Perez, and J.~Bouwen, ``Quantifying the value of 5g
  and edge cloud on qoe for ar/vr,'' in \emph{2020 Twelfth International
  Conference on Quality of Multimedia Experience (QoMEX)}.\hskip 1em plus 0.5em
  minus 0.4em\relax IEEE, 2020, pp. 1--4.

\bibitem{10.1117/12.3013532}
\BIBentryALTinterwordspacing
N.~Rashvand, K.~Witham, G.~Maldonado, V.~Katariya, A.~Sultan, G.~Schirner, and
  H.~Tabkhi, ``{Distributed learning for automatic modulation recognition in
  bandwidth-limited networks},'' in \emph{Signal Processing, Sensor/Information
  Fusion, and Target Recognition XXXIII}, I.~Kadar, E.~P. Blasch, and L.~L.
  Grewe, Eds., vol. 13057, International Society for Optics and
  Photonics.\hskip 1em plus 0.5em minus 0.4em\relax SPIE, 2024, p. 130570X.
  [Online]. Available: \url{https://doi.org/10.1117/12.3013532}
\BIBentrySTDinterwordspacing

\bibitem{aukstakalnis2016practical}
S.~Aukstakalnis, \emph{Practical augmented reality: A guide to the
  technologies, applications, and human factors for AR and VR}.\hskip 1em plus
  0.5em minus 0.4em\relax Addison-Wesley Professional, 2016.

\bibitem{lee2022evaluation}
H.~Lee, J.~Kim, J.-Y. Son, I.~Kim, J.~Noh, Y.-J. Yoon, and M.~Yoon,
  ``Evaluation of eye response using a wearable display with automatic
  interpupillary distance adjustment,'' \emph{Optics Express}, vol.~30, no.~5,
  pp. 8151--8164, 2022.

\bibitem{dincelli2020choose}
E.~Dincelli and I.~Chengalur-Smith, ``Choose your own training adventure:
  designing a gamified seta artefact for improving information security and
  privacy through interactive storytelling,'' \emph{European Journal of
  Information Systems}, vol.~29, no.~6, pp. 669--687, 2020.

\bibitem{jang2019progress}
H.~J. Jang, J.~Y. Lee, J.~Kwak, D.~Lee, J.-H. Park, B.~Lee, and Y.~Y. Noh,
  ``Progress of display performances: Ar, vr, qled, oled, and tft,''
  \emph{Journal of Information Display}, vol.~20, no.~1, pp. 1--8, 2019.

\bibitem{zhan2020augmented}
T.~Zhan, K.~Yin, J.~Xiong, Z.~He, and S.-T. Wu, ``Augmented reality and virtual
  reality displays: perspectives and challenges,'' \emph{Iscience}, vol.~23,
  no.~8, p. 101397, 2020.

\bibitem{xiong2021augmented}
J.~Xiong, E.-L. Hsiang, Z.~He, T.~Zhan, and S.-T. Wu, ``Augmented reality and
  virtual reality displays: emerging technologies and future perspectives,''
  \emph{Light: Science \& Applications}, vol.~10, no.~1, p. 216, 2021.

\bibitem{kim2019foveated}
J.~Kim, Y.~Jeong, M.~Stengel, K.~Aksit, R.~A. Albert, B.~Boudaoud, T.~Greer,
  J.~Kim, W.~Lopes, Z.~Majercik \emph{et~al.}, ``Foveated ar:
  dynamically-foveated augmented reality display.'' \emph{ACM Trans. Graph.},
  vol.~38, no.~4, pp. 99--1, 2019.

\bibitem{li2020distortion}
H.~Li, H.~Li, L.~Xu, and X.~Liu, ``Distortion correction and image registration
  of ultralarge field of view near-eye display device,'' \emph{Applied Optics},
  vol.~59, no.~14, pp. 4422--4431, 2020.

\bibitem{hong202256}
W.~Hong, S.~Park, H.~Kim, and J.~G. Lee, ``56-2: A distraction-free display
  system using embedded asynchronous time warp,'' in \emph{SID Symposium Digest
  of Technical Papers}, vol.~53.\hskip 1em plus 0.5em minus 0.4em\relax Wiley
  Online Library, 2022, pp. 736--739.

\bibitem{iseland2020evaluation}
J.~C. Iseland and L.~Grolleman, ``Evaluation of performance on variable rate
  shading,'' \emph{Applied Optics}, 2020.

\bibitem{trillmichevolution}
F.~Trillmich and F.~Wedel, ``The evolution of collaboration tools to facilitate
  internal collaboration,'' \emph{IEEE Transactions on Visualization and
  Computer Graphics}, 2019.

\bibitem{9598519}
A.~Sun, Y.~Tao, M.-L. Shyu, S.-C. Chen, A.~Blizzard, W.~A. Rothenberg,
  D.~Garcia, and J.~F. Jent, ``Multimodal data integration for interactive and
  realistic avatar simulation in augmented reality,'' in \emph{2021 IEEE 22nd
  International Conference on Information Reuse and Integration for Data
  Science (IRI)}, 2021, pp. 362--369.

\bibitem{gamelin2021point}
G.~Gamelin, A.~Chellali, S.~Cheikh, A.~Ricca, C.~Dumas, and S.~Otmane,
  ``Point-cloud avatars to improve spatial communication in immersive
  collaborative virtual environments,'' \emph{Personal and Ubiquitous
  Computing}, vol.~25, pp. 467--484, 2021.

\bibitem{ye2019enhancing}
S.~Ye, T.~Ying, L.~Zhou, and T.~Wang, ``Enhancing customer trust in
  peer-to-peer accommodation: A “soft” strategy via social presence,''
  \emph{International Journal of Hospitality Management}, vol.~79, pp. 1--10,
  2019.

\bibitem{wen2020photorealistic}
X.~Wen, M.~Wang, C.~Richardt, Z.-Y. Chen, and S.-M. Hu, ``Photorealistic
  audio-driven video portraits,'' \emph{IEEE Transactions on Visualization and
  Computer Graphics}, vol.~26, no.~12, pp. 3457--3466, 2020.

\bibitem{park2019study}
W.~Park, H.~Heo, S.~Park, and J.~Kim, ``A study on the presence of immersive
  user interface in collaborative virtual environments application,''
  \emph{Symmetry}, vol.~11, no.~4, p. 476, 2019.

\bibitem{derby2021challenges}
J.~L. Derby and B.~S. Chaparro, ``The challenges of evaluating the usability of
  augmented reality (ar),'' in \emph{Proceedings of the Human Factors and
  Ergonomics Society Annual Meeting}, vol.~65.\hskip 1em plus 0.5em minus
  0.4em\relax SAGE Publications Sage CA: Los Angeles, CA, 2021, pp. 994--998.

\bibitem{kamarulzaman2020comparative}
N.~A. Kamarulzaman, N.~Fabil, Z.~M. Zaki, and R.~Ismail, ``Comparative study of
  icon design for mobile application,'' in \emph{Journal of Physics: Conference
  Series}, vol. 1551.\hskip 1em plus 0.5em minus 0.4em\relax IOP Publishing,
  2020, p. 012007.

\bibitem{whitlock2020authar}
M.~Whitlock, G.~Fitzmaurice, T.~Grossman, and J.~Matejka, ``Authar: concurrent
  authoring of tutorials for ar assembly guidance,'' in \emph{Graphics
  Interface 2020}, 2020.

\bibitem{9975360}
D.~B. Markant, M.~Rogha, A.~Karduni, R.~Wesslen, and W.~Dou, ``Can data
  visualizations change minds? identifying mechanisms of elaborative thinking
  and persuasion,'' in \emph{2022 IEEE Workshop on Visualization for Social
  Good (VIS4Good)}, 2022, pp. 1--5.

\bibitem{make6010010}
\BIBentryALTinterwordspacing
M.~Fallahian, M.~Dorodchi, and K.~Kreth, ``Gan-based tabular data generator for
  constructing synopsis in approximate query processing: Challenges and
  solutions,'' \emph{Machine Learning and Knowledge Extraction}, vol.~6, no.~1,
  pp. 171--198, 2024. [Online]. Available:
  \url{https://www.mdpi.com/2504-4990/6/1/10}
\BIBentrySTDinterwordspacing

\bibitem{4637362}
F.~Zhou, H.~B.-L. Duh, and M.~Billinghurst, ``Trends in augmented reality
  tracking, interaction and display: A review of ten years of ismar,'' in
  \emph{2008 7th IEEE/ACM International Symposium on Mixed and Augmented
  Reality}, 2008, pp. 193--202.

\bibitem{9146929}
Y.~Dong, L.~Ji, S.~Wang, P.~Gong, J.~Yue, R.~Shen, C.~Chen, and Y.~Zhang,
  ``Accurate 6dof pose tracking for texture-less objects,'' \emph{IEEE
  Transactions on Circuits and Systems for Video Technology}, vol.~31, no.~5,
  pp. 1834--1848, 2021.

\bibitem{10.1145/3126594.3126664}
\BIBentryALTinterwordspacing
P.-C. Wu, R.~Wang, K.~Kin, C.~Twigg, S.~Han, M.-H. Yang, and S.-Y. Chien,
  ``Dodecapen: Accurate 6dof tracking of a passive stylus,'' in
  \emph{Proceedings of the 30th Annual ACM Symposium on User Interface Software
  and Technology}, ser. UIST '17.\hskip 1em plus 0.5em minus 0.4em\relax New
  York, NY, USA: Association for Computing Machinery, 2017, p. 365–374.
  [Online]. Available: \url{https://doi.org/10.1145/3126594.3126664}
\BIBentrySTDinterwordspacing

\bibitem{eger2020measuring}
D.~Eger~Passos and B.~Jung, ``Measuring the accuracy of inside-out tracking in
  xr devices using a high-precision robotic arm,'' in \emph{HCI International
  2020-Posters: 22nd International Conference, HCII 2020, Copenhagen, Denmark,
  July 19--24, 2020, Proceedings, Part I}.\hskip 1em plus 0.5em minus
  0.4em\relax Springer, 2020, pp. 19--26.

\bibitem{sorriento2019optical}
A.~Sorriento, M.~B. Porfido, S.~Mazzoleni, G.~Calvosa, M.~Tenucci, G.~Ciuti,
  and P.~Dario, ``Optical and electromagnetic tracking systems for biomedical
  applications: A critical review on potentialities and limitations,''
  \emph{IEEE reviews in biomedical engineering}, vol.~13, pp. 212--232, 2019.

\bibitem{semwal2022pattern}
V.~B. Semwal, N.~Gaud, P.~Lalwani, V.~Bijalwan, and A.~K. Alok, ``Pattern
  identification of different human joints for different human walking styles
  using inertial measurement unit (imu) sensor,'' \emph{Artificial Intelligence
  Review}, vol.~55, no.~2, pp. 1149--1169, 2022.

\bibitem{9085357}
S.~Su, H.~Dai, S.~Cheng, P.~Lin, C.~Hu, and B.~Lv, ``A robust magnetic tracking
  approach based on graph optimization,'' \emph{IEEE Transactions on
  Instrumentation and Measurement}, vol.~69, no.~10, pp. 7933--7940, 2020.

\bibitem{kodeboyina2016low}
S.~M. Kodeboyina and K.~Varghese, ``Low cost augmented reality framework for
  construction applications,'' in \emph{ISARC. Proceedings of the International
  Symposium on Automation and Robotics in Construction}, vol.~33.\hskip 1em
  plus 0.5em minus 0.4em\relax IAARC Publications, 2016, p.~1.

\bibitem{parham2019creating}
G.~Parham, E.~G. Bing, A.~Cuevas, B.~Fisher, J.~Skinner, M.~Mwanahamuntu, and
  R.~Sullivan, ``Creating a low-cost virtual reality surgical simulation to
  increase surgical oncology capacity and capability,''
  \emph{ecancermedicalscience}, vol.~13, 2019.

\bibitem{sorko2019potentials}
S.~R. Sorko and M.~Brunnhofer, ``Potentials of augmented reality in training,''
  \emph{Procedia Manufacturing}, vol.~31, pp. 85--90, 2019.

\bibitem{8942324}
V.~Chheang, P.~Saalfeld, T.~Huber, F.~Huettl, W.~Kneist, B.~Preim, and
  C.~Hansen, ``Collaborative virtual reality for laparoscopic liver surgery
  training,'' in \emph{2019 IEEE International Conference on Artificial
  Intelligence and Virtual Reality (AIVR)}, 2019, pp. 1--17.

\bibitem{kapp2022design}
K.~Kapp, M.~Siv{\'e}n, P.~Laur{\'e}n, S.~Virtanen, N.~Katajavuori, and
  I.~S{\"o}dervik, ``Design and usability testing of an augmented reality (ar)
  environment in pharmacy education—presenting a pilot study on comparison
  between ar smart glasses and a mobile device in a laboratory course,''
  \emph{Education Sciences}, vol.~12, no.~12, p. 854, 2022.

\bibitem{9619037}
S.~Kirakosian, G.~Daskalogrigorakis, E.~Maravelakis, and K.~Mania,
  ``Near-contact person-to-3d character dance training: Comparing ar and vr for
  interactive entertainment,'' in \emph{2021 IEEE Conference on Games (CoG)},
  2021, pp. 1--5.

\bibitem{li2021application}
B.~Li and X.~Xu, ``Application of artificial intelligence in basketball
  sport,'' \emph{Journal of Education, Health and Sport}, vol.~11, no.~7, pp.
  54--67, 2021.

\bibitem{10.1145/3411764.3445649}
\BIBentryALTinterwordspacing
T.~Lin, R.~Singh, Y.~Yang, C.~Nobre, J.~Beyer, M.~A. Smith, and H.~Pfister,
  ``Towards an understanding of situated ar visualization for basketball
  free-throw training,'' in \emph{Proceedings of the 2021 CHI Conference on
  Human Factors in Computing Systems}, ser. CHI '21.\hskip 1em plus 0.5em minus
  0.4em\relax New York, NY, USA: Association for Computing Machinery, 2021.
  [Online]. Available: \url{https://doi.org/10.1145/3411764.3445649}
\BIBentrySTDinterwordspacing

\bibitem{thomason2021metahealth}
J.~Thomason, ``Metahealth-how will the metaverse change health care?''
  \emph{Journal of Metaverse}, vol.~1, no.~1, pp. 13--16, 2021.

\bibitem{horrell2021telemedicine}
L.~N. Horrell, S.~Hayes, L.~B. Herbert, K.~MacTurk, L.~Lawhon, C.~G. Valle, and
  A.~Bhowmick, ``Telemedicine use and health-related concerns of patients with
  chronic conditions during covid-19: survey of members of online health
  communities,'' \emph{Journal of Medical Internet Research}, vol.~23, no.~2,
  p. e23795, 2021.

\bibitem{9222346}
B.~Lee, X.~Hu, M.~Cordeil, A.~Prouzeau, B.~Jenny, and T.~Dwyer, ``Shared
  surfaces and spaces: Collaborative data visualisation in a co-located
  immersive environment,'' \emph{IEEE Transactions on Visualization and
  Computer Graphics}, vol.~27, no.~2, pp. 1171--1181, 2021.

\bibitem{pedram2020examining}
S.~Pedram, S.~Palmisano, P.~Perez, R.~Mursic, and M.~Farrelly, ``Examining the
  potential of virtual reality to deliver remote rehabilitation,''
  \emph{Computers in Human Behavior}, vol. 105, p. 106223, 2020.

\bibitem{eckert2019augmented}
M.~Eckert, J.~S. Volmerg, C.~M. Friedrich \emph{et~al.}, ``Augmented reality in
  medicine: systematic and bibliographic review,'' \emph{JMIR mHealth and
  uHealth}, vol.~7, no.~4, p. e10967, 2019.

\bibitem{guo2020applications}
Z.~Guo, D.~Zhou, Q.~Zhou, X.~Zhang, J.~Geng, S.~Zeng, C.~Lv, and A.~Hao,
  ``Applications of virtual reality in maintenance during the industrial
  product lifecycle: A systematic review,'' \emph{Journal of Manufacturing
  Systems}, vol.~56, pp. 525--538, 2020.

\bibitem{8713454}
S.~Altarteer and V.~Charissis, ``Technology acceptance model for 3d virtual
  reality system in luxury brands online stores,'' \emph{IEEE Access}, vol.~7,
  pp. 64\,053--64\,062, 2019.

\bibitem{qin2021virtual}
H.~Qin, D.~A. Peak, and V.~Prybutok, ``A virtual market in your pocket: How
  does mobile augmented reality (mar) influence consumer decision making?''
  \emph{Journal of Retailing and Consumer Services}, vol.~58, p. 102337, 2021.

\bibitem{bates1992virtual}
J.~Bates, ``Virtual reality, art, and entertainment,'' \emph{Presence:
  Teleoperators \& Virtual Environments}, vol.~1, no.~1, pp. 133--138, 1992.

\bibitem{pallavicini2019gaming}
F.~Pallavicini, A.~Pepe, and M.~E. Minissi, ``Gaming in virtual reality: What
  changes in terms of usability, emotional response and sense of presence
  compared to non-immersive video games?'' \emph{Simulation \& Gaming},
  vol.~50, no.~2, pp. 136--159, 2019.

\bibitem{shin2019does}
D.~Shin, ``How does immersion work in augmented reality games? a user-centric
  view of immersion and engagement,'' \emph{Information, Communication \&
  Society}, vol.~22, no.~9, pp. 1212--1229, 2019.

\bibitem{9223669}
P.~Reipschlager, T.~Flemisch, and R.~Dachselt, ``Personal augmented reality for
  information visualization on large interactive displays,'' \emph{IEEE
  Transactions on Visualization and Computer Graphics}, vol.~27, no.~2, pp.
  1182--1192, 2021.

\end{thebibliography}

\end{document}